\documentclass[aps,twocolumn,superscriptaddress,amsmath]{revtex4-1}
\usepackage[pdftex]{graphicx,color,hyperref}
\usepackage{soul}
\usepackage{amsfonts}
\usepackage{amssymb}
\usepackage{amsmath}
\usepackage{amsthm}
\usepackage{bm}
\usepackage{braket}
\usepackage{enumerate}

\begin{document}

\title{Quenched vs Annealed: Glassiness from SK to SYK}

\author{C. L. Baldwin}
\affiliation{National Institute of Standards and Technology, Gaithersburg, MD 20899, USA}
\affiliation{Joint Quantum Institute, University of Maryland, College Park, MD 20742, USA}

\author{B. Swingle}
\affiliation{Condensed Matter Theory Center, University of Maryland, College Park, MD 20742, USA}
\affiliation{Maryland Center for Fundamental Physics, University of Maryland, College Park, MD 20742, USA}
\affiliation{Joint Institute for Quantum Information and Computer Science, University of Maryland, College Park, MD 20742, USA}
\affiliation{Department of Physics, University of Maryland, College Park, MD 20742, USA}

\date{\today}

\begin{abstract}

We show that any SYK-like model with finite-body interactions among \textit{local} degrees of freedom, e.g., bosons or spins, has a fundamental difference from the standard fermionic model: the former fails to be described by an annealed free energy at low temperature.
In this respect, such models more closely resemble spin glasses.
We demonstrate this by two means: first, a general theorem proving that the annealed free energy is divergent at low temperature in any model with a tensor product Hilbert space; and second, a replica treatment of two prominent examples which exhibit phase transitions from an ``annealed'' phase to a ``non-annealed`` phase as a function of temperature.
We further show that this effect appears only at $O(N)$'th order in a $1/N$ expansion, even though lower-order terms misleadingly seem to converge.
Our results prove that the non-bosonic nature of the particles in SYK is an essential ingredient for its physics, highlight connections between local models and spin glasses, and raise important questions as to the role of fermions and/or glassiness in holography.

\end{abstract}

\maketitle

\tableofcontents

\section{Introduction} \label{sec:introduction}

There has recently been tremendous interest in the Sachdev-Ye-Kitaev (SYK) model of interacting Majorana fermions~\cite{Sachdev1993Gapless,Georges2000Mean,Georges2001Quantum,kitaev,Polchinski_2016,ms,kamenev,Garc_a_Garc_a_2016}. This is largely because the low-energy limit of SYK provides a tractable example of holography: a duality between a quantum system without gravity and a quantum system with gravity in an emergent dynamical spacetime~\cite{Maldacena:1997re,Gubser_1998,witten1998anti}. It has become a valuable toy model in studies on the chaotic nature of black holes, the black hole information problem, and much more. It has also given rise to models of strongly correlated electronic systems~\cite{Gu_2017,alex2019syk,Guo_2019,Can_2019a,Kim_2019,Can_2019b,dai2018global,Chowdhury_2018,Ben_Zion_2018}, and has inspired proposals for experimental realizations~\cite{Danshita_2017,Pikulin_2017,Garc_a_lvarez_2017,Babbush_2019}.

What the SYK model does not exhibit is spin glass physics~\cite{GurAri2018Does,Wang2018On,Arefeva2019Replica}.
This is surprising, because the SYK Hamiltonian bears a striking similarity to the quintessential mean-field models of spin glass theory~\cite{Sherrington1975Solvable,Derrida1980Random,Bray1980Replica}.
Both the SYK and spin glass models are defined by random-strength interactions among all degrees of freedom.
Here we show that the essential difference is the fermionic nature of the particles in SYK: any model with strictly local degrees of freedom will share much more in common with spin glasses.

This result is relevant because interest in SYK physics has spread to generalizations of the original model.
To name a few: including multiple flavors of fermions~\cite{Gross2017Generalization}, using bosonic particles~\cite{Georges2000Mean,Biroli2002Out,Fu2016Numerical}, using spins~\cite{Berkooz2018Chord,Berkooz2019Towards}, forming lattices of SYK models~\cite{Gu2017Local,Song2017Strongly}, and introducing supersymmetry~\cite{Fu2017Supersymmetric,Li2017Supersymmetric}.
With the analysis presented in this paper, we are able to immediately identify large classes of such models in which the potential for glassiness must be carefully addressed.

On the spin glass side, all-to-all disordered models have featured prominently for decades.
Sherrington and Kirkpatrick first introduced a system of Ising spins with infinite-range random interactions which exhibits an intricate spin glass phase~\cite{Sherrington1975Solvable,Thouless1977Solution,Parisi1979Infinite}.
The model has been extended in numerous directions, both classical and quantum, many of which are central to the field in their own right: $p$-body interactions~\cite{Derrida1980Random,Gardner1985Spin,Kirkpatrick1987Dynamics}, spherical spins~\cite{Crisanti1992Spherical,Crisanti1993Spherical,Cugliandolo1993Analytical}, Potts spins~\cite{Gross1985Mean,Kirkpatrick1987Stable}, Heisenberg interactions~\cite{Bray1980Replica,Sommers1981Theory,Sachdev1993Gapless,Georges2000Mean}, and transverse fields~\cite{Usadel1987Quantum,Goldschmidt1990Solvable,Nieuwenhuizen1998Quantum}, among many others.

These variants all share certain phenomena which unite them as spin glasses.
As one lowers the temperature, the system first undergoes a ``dynamical'' transition at temperature $T_d$, below which dynamical correlation functions never fully decay.
The system experiences a further ``static'' transition at a potentially lower $T_s$, below which one can detect frozen magnetization patterns in the equilibrium Gibbs distribution.
Certain systems undergo a third ``Gardner'' transition at an even lower $T_g$, below which the magnetization patterns become more complex, with sub-patterns and so on.
For pedagogical expositions of the physics, see Refs.~\cite{Mezard1987,Castellani2005Spin,Mezard2009}.

All indications are that the SYK model does not show any such behavior~\cite{Sachdev1993Gapless,Bagrets2016Sachdev,GurAri2018Does,Wang2018On,Arefeva2019Replica}.
This raises multiple questions, chief among which is simply: which generalizations of the SYK model \textit{do} have spin glass phases?
Presumably such glassiness would rule out any connection to quantum gravity (although that is itself an important open question).
It has long been known that the bosonic variant of SYK is a spin glass~\cite{Georges2000Mean,Georges2001Quantum,Biroli2002Out}, yet this is merely one model out of the multitude which could arise.
A recent numerical study on small systems found evidence suggesting that the hard-core bosonic variant is a spin glass as well~\cite{Fu2016Numerical}.
Beyond this, the question has remained unexplored.
There has been no general framework for understanding when all-to-all disordered systems behave as spin glasses rather than SYK.

This paper aims to fill that gap.
On a technical level, generic models can be analyzed in two ways, and we address both.
The first relies on the replica formalism: one expresses the moments of the partition function as a path integral and uses standard mean-field techniques to obtain the free energy~\cite{Mezard1987,Castellani2005Spin,Mezard2009}.
One can circumvent replicas by making the ``annealed'' approximation, namely replacing the partition function by its first moment at the outset.
The second approach is to organize the diagrammatic expansion of the propagator in powers of system size $N$.
One averages each term over the disorder and finds that a summable set of diagrams (the so-called ``melons'' in SYK) gives the leading-in-$N$ contribution.
This ultimately gives the same results as the annealed approximation.
Even though the annealed approximation appears to be correct for the SYK model, it is in general extremely unreliable at low temperature.
Indeed, breakdown of the annealed approximation is often what signals entry into a spin glass phase.
We will be studying this breakdown and its consequences in generic all-to-all disordered systems.

In Sec.~\ref{sec:models}, we introduce our notation and the specific models which will serve as our examples.
In Sec.~\ref{sec:general_breakdown}, we prove that the annealed approximation cannot hold at low temperature in \textit{any} model for which the Hilbert space is a tensor product.
This includes bosons (soft- \& hard-core), spins, distinguishable particles, etc.
It shows that all such models are fundamentally different from SYK.
In Sec.~\ref{sec:specific_analysis}, we then give a more detailed and transparent analysis of the hard-core bosonic and quantum $p$-spin models.
Despite the models not being fully solvable, we show that each undergoes a transition from an annealed phase at high temperature to a non-annealed phase at low temperature.
Lastly, in Sec.~\ref{sec:N_expansion}, we use a concrete example to demonstrate the difficulty in obtaining such results through a $1/N$ expansion.

\section{Models and definitions} \label{sec:models}

Here we define the models of interest, starting with the original SYK model and then introducing various modifications.
All of the models discussed here are in fact ensembles of Hamiltonians given by Gaussian random couplings.
We also give a very brief description of the replica method.
More detailed accounts can be found in the references.

\begin{itemize}
    \item \textbf{Fermionic models:} The original SYK model is defined using $N$ Majorana (i.e., Hermitian) fermion operators $\hat{\gamma}_i$.
    Note that the Hilbert space of the theory has dimension $2^{N/2}$.
    The Hamiltonian, which has an even integer $q$ as a parameter, is
    \begin{equation} \label{eq:SYK_Hamiltonian}
    H_{\textrm{SYK}} = i^{q/2} \sum_{i_1 < \cdots < i_q} J_{i_1 \cdots i_q} \hat{\gamma}_{i_1} \cdots  \hat{\gamma}_{i_q},
    \end{equation}
    where the couplings $J_{i_1 \cdots i_q}$ are independent Gaussian random variables with mean zero and variance
    \begin{equation} \label{eq:real_coupling_variance}
    \textrm{Var} \big[ J_{i_1 \cdots i_q} \big] = \frac{(q-1)!}{N^{q-1}}.
    \end{equation}
    One can also consider the analogous complex SYK model, where the Majorana operators are replaced by complex fermions $\hat{c}_i$ and $\hat{c}_i^\dag$ ($p \equiv q/2$ of each):
    \begin{equation} \label{eq:cSYK_Hamiltonian}
    H_{\textrm{cSYK}} = \sum_{II'} J_{II'} \hat{c}^\dag_{i_1} \cdots \hat{c}^\dag_{i_p} \hat{c}_{i'_p} \cdots \hat{c}_{i'_1}.
    \end{equation}
    Here and throughout, we use a convenient notation in which the multi-index $I$ represents a set of $p$ indices $i_1 < \cdots < i_p$ arranged in increasing order.
    Thus $H_{\textrm{cSYK}}$ consists of all possible $p$-body interactions. The couplings $J_{II'}$ are again independent Gaussians, but now complex with
    \begin{equation} \label{eq:complex_coupling_variance}
    \textrm{Var} \big[ \textrm{Re} J_{II'} \big] = \textrm{Var} \big[ \textrm{Im} J_{II'} \big] = \frac{(p!)^2}{2 N^{2p-1}},
    \end{equation}
    and such that $J_{I'I} = J_{II'}^*$.
    One reason for considering $H_{\textrm{cSYK}}$ as opposed to $H_{\textrm{SYK}}$ (or vice-versa) is that $H_{\textrm{cSYK}}$ has a conserved particle number.
    \item \textbf{Bosonic models:} The bosonic SYK model simply replaces the fermionic operators $\hat{c}_i$ with bosonic operators $\hat{b}_i$:
    \begin{equation} \label{eq:bSYK_Hamiltonian}
    H_{\text{bSYK}} = \sum_{II'} J_{II'} \hat{b}^\dagger_{i_1} \cdots \hat{b}^\dagger_{i_p} \hat{b}_{i'_1} \cdots \hat{b}_{i'_p}.
    \end{equation}
    The couplings $J_{II'}$ remain exactly as in Eq.~\eqref{eq:complex_coupling_variance}.
    An issue with this definition is that in the grand-canonical ensemble, where the number of particles is unlimited, $H_{\textrm{bSYK}}$ is unbounded from below.
    One could therefore work at fixed particle number, as past works on the bosonic SYK model have done~\cite{Sachdev1993Gapless,Georges2000Mean,Biroli2002Out}, or one could interpret the $\hat{b}_i$ as \textit{hard-core} bosons~\cite{Fu2016Numerical} (i.e., exclude double occupancies on sites).
    Either choice guarantees that the model has a definite ground state.
    We shall do the latter: in addition to being more interesting (in the sense that much less is known about it), the hard-core model has the benefit of having a well-defined grand-canonical ensemble.
    \item \textbf{Spin models:} The quantum $p$-spin model consists of all-to-all $p$-body interactions among spins $\hat{\sigma}_i^{\alpha}$ ($\alpha \in \{ x, y, z \}$):
    \begin{equation} \label{eq:p_spin_Hamiltonian}
    H_p =  \sum_{I A} J_I^A \hat{\sigma}_{i_1}^{\alpha_1} \cdots \hat{\sigma}_{i_p}^{\alpha_p} ,
    \end{equation}
    where $I$ is the same multi-index as before and $A = \{ \alpha_1, \cdots, \alpha_p \}$.
    We will use spin-1/2, but our results apply to any spin.
    The couplings $J_I^A$ are real Gaussians with variance
    \begin{equation} \label{eq:p_spin_coupling_variance}
    \textrm{Var} \big[ J_I^A \big] = \frac{p!}{6 (3N)^{p-1}}.
    \end{equation}
    Connections between this spin model and SYK have recently been explored in Refs.~\cite{Berkooz2018Chord,Berkooz2019Towards}.
\end{itemize}

It should be stressed that our conclusions are in no way restricted to these models.
We focus on those listed here solely for the sake of concreteness and current relevance.

Regardless of the model, one is always faced with the question of how to treat the random couplings (the ``disorder'').
Assuming the ultimate goal is to calculate the statistics of physical observables, an important quantity is the ``quenched'' free energy:
\begin{equation} \label{eq:quenched_free_energy_definition}
f(\beta) \equiv -\lim_{N \rightarrow \infty} \frac{1}{N \beta} \mathbb{E} \Big[ \ln{\textrm{Tr} e^{-\beta H}} \Big] ,
\end{equation}
where $\mathbb{E} [ \; \cdot \; ]$ denotes the average over random couplings and $\textrm{Tr} [ \; \cdot \; ]$ is the usual sum over states.
Derivatives of $f(\beta)$ clearly give the disorder-averaged values of observables, exactly as the free energy does in non-random systems.

$f(\beta)$ is extremely difficult to evaluate, even for classical systems.
The replica method is one of the few ways to make analytic progress.
It is based on the identity
\begin{equation} \label{eq:replica_method_identity}
\mathbb{E} \Big[ \ln{\textrm{Tr} e^{-\beta H}} \Big] = \lim_{n \rightarrow 0} \frac{1}{n} \ln{\mathbb{E} \Big[ \left( \textrm{Tr} e^{-\beta H} \right) ^n \Big] }.
\end{equation}
One evaluates the average on the right-hand side for integer $n$, interpreting $(\textrm{Tr} e^{-\beta H})^n$ as the partition function for $n$ uncoupled ``replicas'' of the system, i.e.,
\begin{equation} \label{eq:replica_interpretation}
\begin{aligned}
&\left( \sum_{\Psi} \langle \Psi | e^{-\beta H} | \Psi \rangle \right) ^n \\
& \qquad = \sum_{\Psi_{1} \cdots \Psi_{n}} \langle \Psi_{1} | e^{-\beta H} | \Psi_{1} \rangle \cdots \langle \Psi_{n} | e^{-\beta H} | \Psi_{n} \rangle \\
& \qquad = \sum_{\Psi_{1} \cdots \Psi_{n}} \langle \Psi_{1}, \cdots, \Psi_{n} | e^{-\beta H_{1} - \cdots - \beta H_{n}} | \Psi_{1}, \cdots, \Psi_{n} \rangle ,
\end{aligned}
\end{equation}
where $\{ \Psi \}$ is a complete set of states.
In the first line, the operators and states live in the original Hilbert space $\mathcal{H}$, whereas in the second line, they live in the product space $\mathcal{H}^{\otimes n}$.
Assuming one can obtain an analytic expression for the disorder average of Eq.~\eqref{eq:replica_interpretation}, one then pretends that $n$ is an arbitrary real number and takes the $n \rightarrow 0$ limit.
This technique is clearly not rigorous.
It has nonetheless been tremendously successful in the study of disordered systems~\cite{Mezard1987,Castellani2005Spin,Mezard2009}.

A drastic but useful approximation which avoids replicas entirely is to interchange the disorder average and logarithm in the definition of the quenched free energy (and then take the average inside the trace).
This gives the ``annealed'' free energy:
\begin{equation} \label{eq:annealed_free_energy_definition}
f^{(\textrm{ann})}(\beta) = -\lim_{N \rightarrow \infty} \frac{1}{N \beta} \ln{\mathbb{E} \Big[ \textrm{Tr} e^{-\beta H_g} \Big] }.
\end{equation}
Note that derivatives of $f^{(\textrm{ann})}(\beta)$ do \textit{not} correspond to physical quantities.
One often finds that $f \sim f^{(\textrm{ann})}$ at high temperature but that $f^{(\textrm{ann})}$ gives patently incorrect results at low temperature (see Sec.~\ref{sec:general_breakdown}).
The SYK model seems to be the only known non-trivial counterexample.

As an aside, the terminology ``quenched'' versus ``annealed'' comes from metallurgy, and refers to whether fluctuations in the disorder (accounted for by $\mathbb{E} [ \; \cdot \; ]$) are treated on the same footing as thermal fluctuations in the degrees of freedom (accounted for by the trace).
Eq.~\eqref{eq:quenched_free_energy_definition} treats the disorder as fixed when computing observables and only afterwards averages over disorder, whereas Eq.~\eqref{eq:annealed_free_energy_definition} sums over fluctuations in both simultaneously.

\section{Breakdown of the annealed approximation in tensor product models} \label{sec:general_breakdown}

Here we prove a general result: the annealed free energy cannot be correct at low temperature for any all-to-all model with a tensor product structure.
Specifically, consider any $N$-particle Hamiltonian of the form
\begin{equation} \label{eq:generic_p_body_Hamiltonian}
H_g = \sum_{IA} J_I^A \hat{O}_{i_1}^{\alpha_1} \cdots \hat{O}_{i_p}^{\alpha_p},
\end{equation}
where $I$ denotes sets of $p$ particles and $A$ denotes sets of $p$ indices from some group of size $k$, and $J_I^A$ is Gaussian with
\begin{equation} \label{eq:generic_coupling_var}
\textrm{Var} \big[ J_I^A \big] = \frac{N}{2} \frac{p!}{(kN)^p}.
\end{equation}
We shall take the operators $O_i^{\alpha}$ to be Hermitian, but models such as complex SYK involving non-Hermitian operators can be treated in the exact same manner.
The only restriction we place on the operators is that they obey a tensor product structure: the Hilbert space $\mathcal{H}$ is a tensor product $\mathcal{H}_1 \otimes \cdots \otimes \mathcal{H}_N$ and $\hat{O}_i^{\alpha}$ is shorthand for $1_1 \otimes \cdots \otimes \hat{O}_i^{\alpha} \otimes \cdots \otimes 1_N$.
The quenched and annealed free energies are, respectively,
\begin{equation} \label{eq:generic_free_energies}
\begin{aligned}
f(\beta) =& \, -\lim_{N \rightarrow \infty} \frac{1}{N \beta} \mathbb{E} \Big[ \ln{\textrm{Tr} e^{-\beta H_g}} \Big] , \\
f^{(\textrm{ann})}(\beta) =& \, -\lim_{N \rightarrow \infty} \frac{1}{N \beta} \ln{\mathbb{E} \Big[ \textrm{Tr} e^{-\beta H_g} \Big] }.
\end{aligned}
\end{equation}
We prove that there is a finite $\beta^*$ such that for $\beta > \beta^*$,
\begin{equation} \label{eq:generic_breakdown_result}
f(\beta) \neq f^{(\textrm{ann})}(\beta).
\end{equation}

\subsection{Warm-up}

Let us first consider a classical model, for which the annealed free energy is easily computed.
The Sherrington-Kirkpatrick (SK) model mentioned in the introduction is
\begin{equation} \label{eq:SK_model}
H_{\textrm{SK}} = \sum_{i < j} J_{ij} \sigma_i^z \sigma_j^z,
\end{equation}
with Ising spins $\sigma_i^z$ and $\textrm{Var}[J_{ij}] = 1/N$.
A simple calculation gives
\begin{equation} \label{eq:SK_average_partition}
\mathbb{E} \Big[ \textrm{Tr} e^{-\beta H_{\textrm{SK}}} \Big] = 2^N \prod_{i < j} e^{\frac{\beta^2}{2N}},
\end{equation}
and thus
\begin{equation} \label{eq:SK_annealed_free_energy}
f^{(\textrm{ann})}(\beta) = -\frac{1}{\beta} \ln{2} - \frac{\beta}{4}.
\end{equation}
Yet if Eq.~\eqref{eq:SK_annealed_free_energy} were the correct expression for the \textit{average} free energy, then the average energy per spin would be $\epsilon = -\beta / 2$ and the average entropy would be $s(\epsilon) = \ln{2} - \epsilon^2$.
This cannot be, since the entropy in a discrete configuration space is non-negative: the number of configurations $\Omega(\epsilon)$ within a small energy window around $\epsilon$ is a non-negative integer, thus $\lim_{N \rightarrow \infty} N^{-1} \ln{\Omega(\epsilon)}$ is either $-\infty$ or non-negative.
The annealed free energy of the SK model must be invalid for $\epsilon < -\sqrt{\ln{2}}$, i.e., $\beta > 2 \sqrt{\ln{2}}$.

\subsection{Generic tensor product models}

\begin{figure}[t]
\centering
\includegraphics[width=1.0\columnwidth]{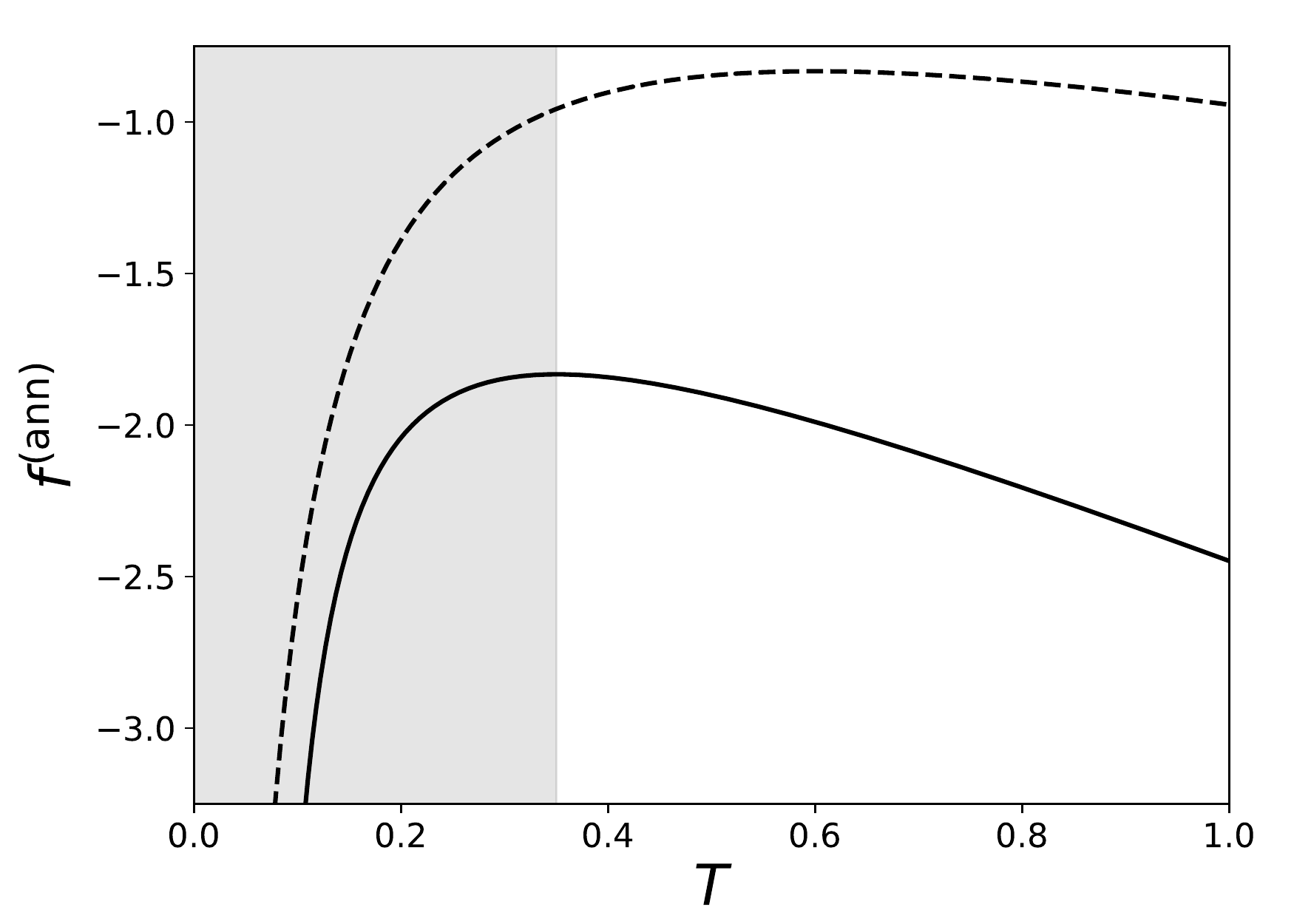}
\caption{Sketch of a bound on the annealed free energy (dashed) which might result from the present analysis, as compared to the exact annealed free energy (solid). The curves shown are merely cartoons, not results for any specific model. The important feature is that the bound diverges to $-\infty$ as $T \rightarrow 0$, forcing the exact curve to do so as well. The shaded region is where $s^{(\textrm{ann})} \equiv - \partial f^{(\textrm{ann})} / \partial T < 0$ and the annealed approximation must be invalid.}
\label{fig:annealed_bound}
\end{figure}

The statement that the entropy must be non-negative applies equally well to quantum systems: simply replace the word ``configurations'' by ``energy eigenstates''.
For any Hamiltonian $H_g$ of the form in Eq.~\eqref{eq:generic_p_body_Hamiltonian}, we give an upper bound to $f^{(\textrm{ann})}$ which diverges to $-\infty$ as $T \equiv 1/\beta \rightarrow 0$.
It follows that the annealed entropy, being $-\partial f^{(\textrm{ann})} / \partial T$, must diverge to $-\infty$ as $T \rightarrow 0$, and thus cannot be correct below a certain temperature. See Fig.~\ref{fig:annealed_bound} for a sketch of the situation.

Since the various $\hat{O}^{\alpha}$ may not commute for different $\alpha$, we cannot directly evaluate the annealed free energy as for the SK model.
Yet we always have Jensen's inequality:
\begin{equation} \label{eq:exponential_Jensen_inequality}
\langle \Psi | e^{-\beta H_g} | \Psi \rangle \geq e^{-\beta \langle \Psi | H_g | \Psi \rangle},
\end{equation}
for any quantum state $| \Psi \rangle$.
Summing Eq.~\eqref{eq:exponential_Jensen_inequality} over a complete set of states, averaging over disorder, and taking the logarithm, we find that
\begin{equation} \label{eq:quantum_classical_inequality}
\begin{aligned}
f^{(\textrm{ann})} \leq & -\frac{1}{N \beta} \ln{\sum_{\Psi} \mathbb{E} \Big[ e^{-\beta \sum_{IA} J_I^A \langle \Psi | \hat{O}_{i_1}^{\alpha_1} \cdots \hat{O}_{i_p}^{\alpha_p} | \Psi \rangle } \Big] } \\
=& -\frac{1}{N \beta} \ln{\sum_{\Psi} e^{\frac{N \beta^2}{4} \frac{p!}{(kN)^p} \sum_{IA} \langle \Psi | \hat{O}_{i_1}^{\alpha_1} \cdots \hat{O}_{i_p}^{\alpha_p} | \Psi \rangle ^2}}.
\end{aligned}
\end{equation}
Note that Eq.~\eqref{eq:quantum_classical_inequality} holds for any basis $| \Psi \rangle$ used on the right-hand side.

The tensor product structure allows us to use a product basis, i.e.,
\begin{equation} \label{eq:generic_tensor_factoring}
\langle \Psi | \hat{O}_{i_1}^{\alpha_1} \cdots \hat{O}_{i_p}^{\alpha_p} | \Psi \rangle = \langle \psi_{i_1} | \hat{O}_{i_1}^{\alpha_1} | \psi_{i_1} \rangle \cdots \langle \psi_{i_p} | \hat{O}_{i_p}^{\alpha_p} | \psi_{i_p} \rangle.
\end{equation}
Furthermore, since the operators $\hat{O}^{\alpha}$ are not identically 0, there must be some single-particle state $| \psi_i^* \rangle$ for which
\begin{equation} \label{eq:generic_single_particle_expectation}
\big| \langle \psi_i^* | \hat{O}_i^{\alpha} | \psi_i^* \rangle \big| \equiv \big| O^{\alpha *} \big| > 0,
\end{equation}
at least for some $\alpha$.
Use this $| \psi_i^* \rangle$ as a basis state.
Then 
\begin{widetext}
\begin{equation} \label{eq:generic_single_term_contribution}
\begin{aligned}
\sum_{\Psi} \exp{\left[ \frac{N \beta^2}{4} \frac{p!}{(kN)^p} \sum_{IA} \langle \Psi | \hat{O}_{i_1}^{\alpha_1} \cdots \hat{O}_{i_p}^{\alpha_p} | \Psi \rangle ^2 \right] } >& \, \exp{\left[ \frac{N \beta^2}{4} \frac{p!}{(kN)^p} \sum_{I A} \big| \langle \psi_{i_1}^* | \hat{O}_{i_1}^{\alpha_1} | \psi_{i_1}^* \rangle \big| ^2 \cdots \big| \langle \psi_{i_p}^* | \hat{O}_{i_p}^{\alpha_p} | \psi_{i_p}^* \rangle \big| ^2 \right] } \\
= & \, \exp{ \left[ \frac{N \beta^2}{4 k^p} \big| O^{\alpha *} \big| ^{2p} + \ldots \, \right] },
\end{aligned}
\end{equation}
\end{widetext}
where the omitted terms (coming from $A \neq \{ \alpha, \cdots , \alpha \}$) are positive.
Inserting into Eq.~\eqref{eq:quantum_classical_inequality} gives our final bound:
\begin{equation} \label{eq:generic_final_bound}
f^{(\textrm{ann})} \leq -\frac{\beta}{4k^p} \big| O^{\alpha *} \big| ^{2p}.
\end{equation}
Clearly $f^{(\textrm{ann})} \rightarrow -\infty$ as $\beta \rightarrow \infty$, as claimed.

The divergence of $f^{(\textrm{ann})}$ at low temperature is not limited to Gaussian disorder.
The Gaussian coupling distribution was used only to evaluate the average in Eq.~\eqref{eq:quantum_classical_inequality}, and an analogous bound can be obtained for any other distribution.
For example, suppose each $J_I^A$ has some alternate probability density $P(J)$ for which the mean is zero and the variance is still given by Eq.~\eqref{eq:generic_coupling_var}.
Assume $P(J)$ falls off faster than exponentially for $J^2 \gg \textrm{Var}[J]$, so that we can safely expand inside the average:
\begin{equation} \label{eq:generic_alternate_exponential_average}
\begin{aligned}
& \mathbb{E} \Big[ e^{-\beta \sum_{IA} J_I^A \langle \Psi | \hat{O}_{i_1}^{\alpha_1} \cdots \hat{O}_{i_p}^{\alpha_p} | \Psi \rangle } \Big] \\
& \quad \sim \prod_{IA} \left( 1 + \frac{1}{2} \beta^2 \mathbb{E} \big[ (J_I^A)^2 \big] \langle \Psi | \hat{O}_{i_1}^{\alpha_1} \cdots \hat{O}_{i_p}^{\alpha_p} | \Psi \rangle ^2 \right) \\
& \quad \sim e^{\frac{N \beta^2}{4} \frac{p!}{(kN)^p} \sum_{IA} \langle \Psi | \hat{O}_{i_1}^{\alpha_1} \cdots \hat{O}_{i_p}^{\alpha_p} | \Psi \rangle ^2 },
\end{aligned}
\end{equation}
We can proceed with the proof as before and obtain the same Eq.~\eqref{eq:generic_final_bound}, for any such $P(J)$.

For the sake of concreteness, we next consider some specific models.

\subsection{Example: Quantum p-spin}

A natural basis to use for the quantum $p$-spin Hamiltonian (Eq.~\eqref{eq:p_spin_Hamiltonian}) is the $\hat{\sigma}_i^z$ eigenstates $| \uparrow \rangle$ and $| \downarrow \rangle$.
Both states have expectation values $| \langle \hat{\sigma}_i^z \rangle |^2 = 1/4$, $| \langle \hat{\sigma}_i^x \rangle |^2 = | \langle \hat{\sigma}_i^y \rangle |^2 = 0$.
Thus Eq.~\eqref{eq:quantum_classical_inequality} gives the bound
\begin{equation} \label{eq:p_spin_annealed_bound}
f_p^{(\textrm{ann})} \leq -\frac{1}{\beta} \ln{2} - \frac{\beta}{4} \frac{1}{12^p}.
\end{equation}
The extra term compared to Eq.~\eqref{eq:generic_final_bound} comes from summing over the $2^N$ basis states, which we neglected for simplicity in the general treatment.

\subsection{Example: Hard-core bosonic SYK}

In this case (Eq.~\eqref{eq:bSYK_Hamiltonian}), we have that
\begin{equation} \label{eq:bSYK_initial_annealed_bound}
f_{\textrm{bSYK}}^{(\textrm{ann})} \leq -\lim_{N \rightarrow \infty} \frac{1}{N \beta} \ln{\sum_{\psi_1 \cdots \psi_N} e^{\frac{N \beta^2}{2} \left( \frac{1}{N} \sum_i | \langle \psi_i | \hat{b}_i | \psi_i \rangle | ^2 \right) ^{2p}}}.
\end{equation}
The number eigenstates $| \psi_i \rangle \in \{ | 0 \rangle, | 1 \rangle \}$ do not give useful bounds, but the superpositions $| \psi_i \rangle \in \{ (| 0 \rangle \pm | 1 \rangle) / \sqrt{2} \}$ do:
\begin{equation} \label{eq:bSYK_annealed_bound}
f_{\textrm{bSYK}}^{(\textrm{ann})} \leq -\frac{1}{\beta} \ln{2} - \frac{\beta}{2} \frac{1}{16^p}.
\end{equation}

\subsection{Example: Complex fermionic SYK}

Note that the bound in Eq.~\eqref{eq:quantum_classical_inequality} \textit{always} applies, regardless of whether the Hilbert space is a tensor product or not.
In particular, it holds for the fermionic SYK model (both real and complex), for which the annealed free energy seems to be correct at all temperatures.
It is informative to see how this result is consistent with the bounds obtained through Eq.~\eqref{eq:quantum_classical_inequality}, in contrast to the examples above.

Given the similarity between hard-core bosons and fermions, and given that the states $(| 0 \rangle \pm | 1 \rangle )/\sqrt{2}$ yielded a free energy diverging at low temperature in the former, let us consider the analogous basis in the fermionic Hilbert space:
\begin{equation} \label{eq:cSYK_analogous_state}
\begin{aligned}
| \Psi \rangle \equiv & \; \frac{1}{\sqrt{2^N}} \big( 1 + (-1)^{\psi_1} \hat{c}_1^{\dag} \big) \cdots \big( 1 + (-1)^{\psi_N} \hat{c}_N^{\dag} \big) | 0 \rangle \\
=& \; \frac{1}{\sqrt{2^N}} \sum_{s_1 \cdots s_N} (-1)^{\vec{\psi} \cdot \vec{s}} \prod_{k : \, s_k = 1} \hat{c}_k^{\dag} | 0 \rangle.
\end{aligned}
\end{equation}
Here the state index $\Psi \in \{ 0, 1 \} ^N$ is denoted as a vector $\vec{\psi}$, and similarly for $\vec{s}$.

Starting from Eq.~\eqref{eq:quantum_classical_inequality}, we need to evaluate
\begin{widetext}
\begin{equation} \label{eq:cSYK_matrix_element}
\langle \Psi | \hat{c}_{i_1}^{\dag} \cdots \hat{c}_{i_p}^{\dag} \hat{c}_{j_p} \cdots \hat{c}_{j_1} | \Psi \rangle = \frac{1}{2^N} \sum_{\vec{s}, \vec{s}'} (-1)^{\vec{\psi} \cdot (\vec{s} + \vec{s}')} \langle 0 | \left( \prod_{k : \, s_k = 1} \hat{c}_k \right) \hat{c}_{i_1}^{\dag} \cdots \hat{c}_{i_p}^{\dag} \hat{c}_{j_p} \cdots \hat{c}_{j_1} \left( \prod_{l : \, s'_l = 1} \hat{c}_l^{\dag} \right) | 0 \rangle.
\end{equation}
\end{widetext}
Note that to leading order in $N$, none of the $i_k$ are equal to any of the $j_l$.
A given term vanishes unless $s'_{i_1} = \cdots = s'_{i_p} = 0$, $s'_{j_1} = \cdots = s'_{j_p} = 1$.
Furthermore, we need $s_k = s'_k$ except for $k \in \{ i_1, \cdots, i_p, j_1, \cdots, j_p \}$, in which case $s_k = 1 - s'_k$.
Thus $(-1)^{\vec{\psi} \cdot (\vec{s} + \vec{s'})} = (-1)^{\psi_{i_1} + \cdots + \psi_{i_p} + \psi_{j_p} + \cdots + \psi_{j_1}}$, which can be taken outside the sum.

Additional minus signs come from rearranging the fermion operators.
First note that factors of $\hat{c}_{i_1} \hat{c}_{i_1}^{\dag}$, $\hat{c}_{j_3} \hat{c}_{j_3}^{\dag}$, etc., in which the two matching operators are adjacent, can be treated as the identity: as a pair, they commute past all other operators, and $\hat{c}_i \hat{c}_i^{\dag} | 0 \rangle = | 0 \rangle$.
Thus owing to the initial order of the $\hat{c}_l^{\dag}$ in Eq.~\eqref{eq:cSYK_analogous_state} (the index increases from left to right), one can convince oneself that we obtain a factor of $-1$ for each $k$ less than $i_1$, each $k$ less than $i_2$, each $l$ less than $j_1$, each $l$ less than $j_2$, and so on.
But now suppose that $j_1 \leq i_1 - 2$.
Each choice of $\vec{s}$ can be associated with an $\vec{r}$ according to $s_{j_1 - 1} = 1 - r_{j_1 - 1}$, $s_{j_1 + 1} = 1 - r_{j_1 + 1}$, with all other $s_l = r_l$.
The two vectors give contributions differing by exactly one minus sign, and therefore sum to 0.

The lesson is that Eq.~\eqref{eq:cSYK_matrix_element} evaluates to 0 unless every $i$ \& $j$ index is adjacent to another, e.g., $i_1 = j_1 - 1$ or $i_2 = j_1 + 1$.
Yet this restricts the number of free indices for us to sum over, and we needed all $2p$ to be free in order to obtain an extensive bound on $f^{(\textrm{ann})}$ (a factor of $N^{2p}$ to compensate for $N^{-(2p-1)}$ from the coupling variance).
In the thermodynamic limit, the only bound we obtain in this case is
\begin{equation} \label{eq:cSYK_specific_bound}
f_{\textrm{cSYK}}^{(\textrm{ann})} \leq -\frac{1}{\beta} \ln{2}.
\end{equation}
The right-hand side does not diverge as $\beta \rightarrow \infty$, and $f_{\textrm{cSYK}}^{(\textrm{ann})}$ has the potential to remain correct even at zero temperature.

\section{Replica analysis for specific models} \label{sec:specific_analysis}

Much additional insight comes from considering the replica analysis in detail for specific models.
We shall focus on the hard-core bosonic and $p$-spin models.
Yet keep in mind that even though we limit ourselves to these two for ease of presentation, our analysis is in fact much more general.
It can be applied with minimal modifications to any model which admits a path integral representation, even if local constraints on the fields are required.

Furthermore, the replica analysis allows us to make conclusions about the high-temperature behavior of the models.
Indeed, we show that the hard-core bosonic and $p$-spin models undergo genuine phase transitions: for each, there exists a $\beta_c$ such that the free energy equals $f^{(\textrm{ann})}$ for $\beta < \beta_c$ and does not for $\beta > \beta_c$.
We are able to say this without needing to calculate the precise functional behavior of $f^{(\textrm{ann})}$.

\subsection{Hard-core bosonic SYK} \label{subsec:hard_core_SYK}

The hard-core bosonic SYK model is given by Eq.~\eqref{eq:bSYK_Hamiltonian}, reproduced here:
\begin{equation}
H_{\text{bSYK}} = \sum_{II'} J_{II'} \hat{b}^\dagger_{i_1} \cdots \hat{b}^\dagger_{i_p} \hat{b}_{i'_1} \cdots \hat{b}_{i'_p}. \tag{\ref{eq:bSYK_Hamiltonian}}
\end{equation}
To construct a path integral representation of the partition function, we express each hard-core boson operator $\hat{b}_i$ as a pair of fermions together with a constraint:
\begin{equation} \label{eq:hard_core_representation}
\hat{b}_i = \hat{h}_i^{\dag} \hat{a}_i, \quad \hat{h}_i^{\dag} \hat{h}_i + \hat{a}_i^{\dag} \hat{a}_i = 1.
\end{equation}
The partition function is then
\begin{widetext}
\begin{equation} \label{eq:hSYK_partition_path_integral}
Z_{\textrm{bSYK}} = \int \prod_i \mathcal{D}h_i \mathcal{D}a_i \mathcal{D}\mu_i \, e^{-\sum_i S^{(0)}[h_i, a_i, \mu_i] - \sum_{II'} J_{II'} S_{\textrm{int}}[h_{II'}, a_{II'}]},
\end{equation}
with
\begin{equation} \label{eq:hSYK_partition_single_particle_action}
S^{(0)}[h_i, a_i, \mu_i] \equiv \int_0^1 \textrm{d}\tau \, \Big( h_i(\tau)^* \partial_{\tau} h_i(\tau) + a_i(\tau)^* \partial_{\tau} a_i(\tau) - \mu_i(\tau) \big( h_i(\tau)^* h_i(\tau) + a_i(\tau)^* a_i(\tau) - 1 \big) \Big) ,
\end{equation}
\begin{equation} \label{eq:hSYK_partition_interaction_action}
S_{\textrm{int}}[h_{II'}, a_{II'}] \equiv \beta \int_0^1 \textrm{d}\tau \, a_{i_1}(\tau)^* h_{i_1}(\tau) \cdots a_{i_p}(\tau)^* h_{i_p}(\tau) h_{i'_p}(\tau)^* a_{i'_p}(\tau) \cdots h_{i'_1}(\tau)^* a_{i'_1}(\tau).
\end{equation}
Note that we enforce the constraint by way of a Lagrange multiplier $\mu_i$ on each site.
Also note that we use a slightly non-standard definition of imaginary time: $\tau$ ranges from 0 to 1 for \textit{all} $\beta$.
This will be convenient in what follows.

As described in Sec.~\ref{sec:models}, we now evaluate $\mathbb{E} [Z_{\textrm{bSYK}}^n]$.
To save space, we give the steps of the calculation in Appendix~\ref{app:replicated_action}.
The method is standard, and analogous calculations can be found in, e.g., Refs.~\cite{Goldschmidt1990Solvable,Georges2001Quantum,Castellani2005Spin,Mezard2009}.
The result is a path integral over an order parameter $G_{rr'}(\tau, \tau')$ and Lagrange multiplier $F_{rr'}(\tau, \tau')$:
\begin{equation} \label{eq:hSYK_integrated_path_integral}
\mathbb{E} \big[ Z_{\textrm{bSYK}}^n \big] = \int \prod_{rr'} \mathcal{D}G_{rr'} \mathcal{D}F_{rr'} \, e^{N \Phi_n[G, F] },
\end{equation}
where
\begin{equation} \label{eq:hSYK_derived_order_parameter_action}
\Phi_n[G, F] \equiv \frac{\beta^2}{2} \sum_{rr'} \int_0^1 \textrm{d}\tau \textrm{d}\tau' \Big( G_{rr'}(\tau, \tau')^p G_{r'r}(\tau', \tau)^p - F_{rr'}(\tau, \tau') G_{r'r}(\tau', \tau) \Big) + \ln{\int \prod_r \mathcal{D}h_r \mathcal{D}a_r \mathcal{D}\mu_r e^{-S^{(\textrm{eff})}[h, a, \mu]}},
\end{equation}
\begin{equation} \label{eq:hSYK_single_site_effective_action}
S^{(\textrm{eff})}[h, a, \mu] \equiv \sum_r S^{(0)}[h_r, a_r, \mu_r] - \frac{\beta^2}{2} \sum_{rr'} \int_0^1 \textrm{d}\tau \textrm{d}\tau' F_{rr'}(\tau, \tau') a_r(\tau)^* h_r(\tau) h_{r'}(\tau')^* a_{r'}(\tau').
\end{equation}
\end{widetext}
The indices $r$ and $r'$ denote different replicas: $r, r' \in \{ 1, \cdots, n \}$.

In the thermodynamic limit, Eq.~\eqref{eq:hSYK_integrated_path_integral} is dominated by the saddle-point value, whose location is determined by the equations
\begin{equation} \label{eq:hSYK_saddle_point_G_equation}
F_{rr'}(\tau, \tau') = 2p \, G_{rr'}(\tau, \tau')^p G_{r'r}(\tau', \tau)^{p-1},
\end{equation}
\begin{equation} \label{eq:hSYK_saddle_point_F_equation}
G_{rr'}(\tau, \tau') = \big< h_r(\tau)^* a_r(\tau) a_{r'}(\tau')^* h_{r'}(\tau') \big> _{\textrm{eff}},
\end{equation}
where $\langle \; \cdot \; \rangle _{\textrm{eff}}$ denotes an expectation value using the effective action of Eq.~\eqref{eq:hSYK_single_site_effective_action}.
From the analysis in Appendix~\ref{app:replicated_action}, one can show that in physical terms, the solution $G_{rr'}(\tau, \tau')$ to Eqs.~\eqref{eq:hSYK_saddle_point_G_equation} and~\eqref{eq:hSYK_saddle_point_F_equation} is simply the equilibrium Green's function for the hard-core bosons in the original model:
\begin{equation} \label{eq:G_physical_interpretation}
G_{rr'}(\tau, \tau') = \big< T \hat{b}_{ir}(\tau) \hat{b}_{ir'}(\tau')^{\dag} \big>,
\end{equation}
where $T$ denotes time ordering.

Thus far, all calculations have been exact.
We cannot proceed any further in full generality, since (unlike in the SYK model) the remaining action for $h$ and $a$ is not quadratic.
Nonetheless, we shall use this starting point to both confirm that the annealed free energy diverges at low temperature and show that it is correct at high temperature.

\subsubsection*{Low temperature}

If we set $n = 1$ in Eq.~\eqref{eq:hSYK_integrated_path_integral}, then we in fact have an expression for the annealed free energy:
\begin{equation} \label{eq:annealed_free_energy_replica_expression}
\max_{G,F} \Phi_1[G, F] = -\beta f^{(\textrm{ann})}(\beta).
\end{equation}
In terms of the order parameter $G(\tau - \tau')$ (note the lack of replica indices and that we have assumed time translation invariance), the expression for $\Phi_1$ is
\begin{widetext}
\begin{equation} \label{eq:hSYK_annealed_action}
\begin{aligned}
\Phi_1[G, F] =& \, \frac{\beta^2}{2} \int_0^1 \textrm{d}\tau \Big( G(\tau)^p G(1 - \tau)^p - F(\tau) G(1 - \tau) \Big) \\
& \qquad \qquad + \ln{\int \mathcal{D}h \mathcal{D}a \mathcal{D}\mu \, e^{-S^{(0)}[h, a, \mu] + \frac{\beta^2}{2} \int_0^1 \textrm{d}\tau \textrm{d}\tau' F(\tau - \tau') a(\tau)^* h(\tau) h(\tau')^* a(\tau')}},
\end{aligned}
\end{equation}
and the saddle-point equations are
\begin{equation} \label{eq:hSYK_annealed_G_equation}
F(\tau - \tau') = 2p \, G(\tau - \tau')^p G(\tau' - \tau)^{p-1},
\end{equation}
\begin{equation} \label{eq:hSYK_annealed_F_equation}
G(\tau - \tau') = \big< h(\tau)^* a(\tau) a(\tau')^* h(\tau') \big> _{\textrm{eff}}.
\end{equation}

At low temperature, the maximizer of Eq.~\eqref{eq:hSYK_annealed_action} is static, i.e., independent of $\tau$.
We show this self-consistently.
Given a $\tau$-independent $F$, we can perform a Hubbard-Stratonovich transformation on the remaining path integral:
\begin{equation} \label{eq:hSYK_annealed_static_transformation}
\begin{aligned}
& \int \mathcal{D}h \mathcal{D}a \mathcal{D}\mu \, e^{-S^{(0)}[h, a, \mu] + \frac{\beta^2 F}{2} \int_0^1 \textrm{d}\tau \textrm{d}\tau' a(\tau)^* h(\tau) h(\tau')^* a(\tau')} \\
& \qquad \qquad \qquad = \int \frac{\textrm{d}z \textrm{d}z^*}{2\pi} e^{-\frac{1}{2}|z|^2} \int \mathcal{D}h \mathcal{D}a \mathcal{D}\mu \, e^{-S^{(0)}[h, a, \mu] + \frac{\beta \sqrt{F}}{2} \int_0^1 \textrm{d}\tau \big( z a(\tau)^* h(\tau) + z^* h(\tau)^* a(\tau) \big) }.
\end{aligned}
\end{equation}
The path integral on the second line is precisely that of a single hard-core boson with $z$-dependent Hamiltonian
\begin{equation} \label{eq:hSYK_static_effective_Hamiltonian}
H^{(\textrm{eff})}(z) = -\frac{\sqrt{F}}{2} \Big( z \hat{b}^{\dag} + z^* \hat{b} \Big) .
\end{equation}
We at last have a tractable expression.
Eq.~\eqref{eq:hSYK_annealed_F_equation} becomes
\begin{equation} \label{eq:hSYK_static_annealed_F_equation}
\begin{aligned}
G =& \; \frac{1}{\int \frac{\textrm{d}z \textrm{d}z^*}{2\pi} e^{-\frac{1}{2} |z|^2} \textrm{Tr} \big[ e^{-\beta H^{(\textrm{eff})}(z)} \big] } \int \frac{\textrm{d}z \textrm{d}z^*}{2\pi} e^{-\frac{1}{2} |z|^2} \textrm{Tr} \big[ e^{-\beta (1 - \tau) H^{(\textrm{eff})}(z)} \hat{b} \, e^{-\beta \tau H^{(\textrm{eff})}(z)} \hat{b}^{\dag} \big] \\
=& \; \frac{1}{\int \frac{\textrm{d}z \textrm{d}z^*}{2\pi} e^{-\frac{1}{2} |z|^2} 2 \cosh{\beta \frac{\sqrt{F}}{2} |z|}} \int \frac{\textrm{d}z \textrm{d}z^*}{2\pi} e^{-\frac{1}{2} |z|^2} \cosh{\beta \tau \frac{\sqrt{F}}{2} |z|} \cosh{\beta (1 - \tau) \frac{\sqrt{F}}{2} |z|}.
\end{aligned}
\end{equation}
\end{widetext}
The right-hand side must be independent of $\tau$ in order to be consistent, and this is indeed what happens at low temperature: for $\tau \gg 1/\beta \sim 0$, we have
\begin{equation} \label{eq:hSYK_static_solution}
G \sim \frac{1}{4}, \quad F \sim \frac{2p}{4^{2p-1}}.
\end{equation}
Returning to Eq.~\eqref{eq:hSYK_annealed_action}, the annealed free energy is
\begin{equation} \label{eq:hSYK_annealed_free_energy}
f^{(\textrm{ann})}(\beta) \sim -\frac{\beta}{2} \frac{1}{16^p} + \cdots,
\end{equation}
where the ellipses denote terms subleading in $\beta$.

Of course, $F(\tau)$ is not strictly static at low but non-zero temperature.
Rather, it has a static component $F$ and a correction $\Delta F(\tau)$, where the correction is non-negligible only for $\tau \lesssim 1 / \beta$.
The presence of $\Delta F(\tau)$ does not change the fact that the correlation time is $O(1 / \beta)$, and thus this ansatz for $F(\tau)$ is fully self-consistent.
Furthermore, $\Delta F(\tau)$ only gives subleading corrections to $f^{(\textrm{ann})}$.
The expression shown in Eq.~\eqref{eq:hSYK_annealed_free_energy} is correct to leading order.

Finally, note that while we obtained the annealed free energy by setting $n = 1$ in $\Phi_n[G, F]$ (Eq.~\eqref{eq:hSYK_derived_order_parameter_action}), the same expression results from instead setting all inter-replica order parameters to 0: set $G_{rr'} = F_{rr'} = 0$ for $r \neq r'$, and take the $n \rightarrow 0$ limit as prescribed (see Sec.~\ref{sec:models}).
Since we know that this expression cannot be correct at low temperature, it follows that the true equilibrium value of the order parameter, whatever it may be, cannot be diagonal in replica indices.

The conclusion is that in the hard-core bosonic SYK model, the autocorrelation function $G(\tau)$ develops a static component as $T \rightarrow 0$ which is responsible for the divergent annealed free energy.
This in turn implies that the system must no longer be replica-diagonal.
It also suggests why the fermionic model should behave differently: there, $G(\tau)$ cannot have a non-zero static component because the Fourier transform only has weight on odd multiples of $\pi$.

\subsubsection*{High temperature}

To show that the annealed free energy \textit{is} correct at high temperature, we place a bound on the probability of a random disorder realization having free energy other than $f^{(\textrm{ann})}$.
Write the (random) partition function $Z$ as $\mathcal{Z} \mathbb{E} [Z]$.
Chebyshev's inequality states that
\begin{equation} \label{eq:partition_function_Chebyshev}
\textrm{Pr} \big[ |\mathcal{Z} - 1| > \eta \big] \leq \frac{\textrm{Var} \big[ \mathcal{Z} \big] }{\eta^2}.
\end{equation}
We shall show that for $\beta$ less than a certain value, $\textrm{Var} [\mathcal{Z}_{\textrm{bSYK}}] \rightarrow 0$ as $N \rightarrow \infty$.
Since
\begin{equation} \label{eq:free_energy_quenched_annealed_connection}
f(\beta) = f^{(\textrm{ann})}(\beta) - \frac{1}{N \beta} \ln{\mathcal{Z}(\beta)},
\end{equation}
it follows from Eq.~\eqref{eq:partition_function_Chebyshev} that $f_{\textrm{bSYK}} = f^{(\textrm{ann})}_{\textrm{bSYK}}$ with probability approaching 1 in the thermodynamic limit.

The second moment of $Z_{\textrm{bSYK}}$ is obtained by setting $n = 2$ in Eq.~\eqref{eq:hSYK_integrated_path_integral}.
We have two order parameters: the intra-replica correlator $G_{rr}(\tau, \tau') \equiv G(\tau - \tau')$ ($r \in \{ 1, 2 \}$) and the inter-replica correlator $G_{12}(\tau, \tau') \equiv Q$, which we take to be static.
Thus
\begin{equation} \label{eq:hSYK_second_moment_partition}
\mathbb{E} \big[ Z_{\textrm{bSYK}}^2 \big] = \int \mathcal{D}G \mathcal{D}F \textrm{d}Q \textrm{d}\lambda \, e^{N \Phi_2[G, F; Q, \lambda]},
\end{equation}
with
\begin{widetext}
\begin{equation} \label{eq:hSYK_second_moment_action}
\begin{aligned}
\Phi_2[G, F; Q, \lambda] =& \; \beta^2 \int_0^1 \textrm{d}\tau \, \Big( G(\tau)^p G(1 - \tau)^p - F(\tau) G(1 - \tau) + Q^{2p} - \lambda Q \Big) \\
+& \, \ln{\int \mathcal{D}h_1 \mathcal{D}h_2 \mathcal{D}a_1 \mathcal{D}a_2 \mathcal{D}\mu_1 \mathcal{D}\mu_2 \, e^{-S^{(0)}[h_1, a_1, \mu_1] - S^{(0)}[h_2, a_2, \mu_2]}} \\[-6pt]
& \qquad \qquad \quad \cdot e^{\frac{\beta^2}{2} \int_0^1 \textrm{d}\tau \textrm{d}\tau' F(\tau - \tau') \big( a_1(\tau)^* h_1(\tau) h_1(\tau')^* a_1(\tau') + a_2(\tau)^* h_2(\tau) h_2(\tau')^* a_2(\tau') \big) } \\[-3pt]
& \qquad \qquad \quad \cdot e^{\frac{\beta^2}{2} \lambda \int_0^1 \textrm{d}\tau \textrm{d}\tau' \big( a_1(\tau)^* h_1(\tau) h_2(\tau')^* a_2(\tau') + a_2(\tau)^* h_2(\tau) h_1(\tau')^* a_1(\tau') \big) }.
\end{aligned}
\end{equation}
\end{widetext}
The saddle-point equations are
\begin{equation} \label{eq:hSYK_second_moment_G_equation}
F(\tau - \tau') = 2p \, G(\tau - \tau')^p G(\tau' - \tau)^{p-1},
\end{equation}
\begin{equation} \label{eq:hSYK_second_moment_F_equation}
G(\tau - \tau') = \big< h_r(\tau)^* a_r(\tau) a_r(\tau')^* h_r(\tau') \big> _{\textrm{eff}},
\end{equation}
\begin{equation} \label{eq:hSYK_second_moment_Q_equation}
\lambda = 2p \, Q^{2p-1},
\end{equation}
\begin{equation} \label{eq:hSYK_second_moment_L_equation}
Q = \big< h_1(\tau)^* a_1(\tau) a_2(\tau')^* h_2(\tau') \big> _{\textrm{eff}}.
\end{equation}

We immediately have one solution to the saddle-point equations.
Denote the solution to the \textit{annealed} equations, Eqs.~\eqref{eq:hSYK_annealed_G_equation} and~\eqref{eq:hSYK_annealed_F_equation}, by $G_{\textrm{eq}}(\tau)$ and $F_{\textrm{eq}}(\tau)$.
It is self-consistent to set
\begin{equation} \label{eq:hSYK_second_moment_high_T_solution}
G(\tau) = G_{\textrm{eq}}(\tau), \quad F(\tau) = F_{\textrm{eq}}(\tau), \quad Q = 0, \quad \lambda = 0,
\end{equation}
in Eqs.~\eqref{eq:hSYK_second_moment_G_equation} through~\eqref{eq:hSYK_second_moment_L_equation}.
Note that this is not a trivial statement: it relies on the fact that in the absence of any inter-replica coupling, each replica has a separate $U(1)$ symmetry which ensures $\langle h_r(\tau)^* a_r(\tau) \rangle = 0$.
If the action were to include an explicit $U(1)$-breaking term, then $Q = 0$ would not be a valid solution.

The question is now whether $Q = 0$ is the dominant solution to the saddle-point equations, i.e., that which maximizes $\Phi_2$.
To address this carefully, in Eq.~\eqref{eq:hSYK_second_moment_action}, denote every part of the expression except for $Q^{2p}$ and $\lambda Q$ by $A_2[G, F; \lambda]$.
This lets us write the second moment of $Z_{\textrm{bSYK}}$ as follows:
\begin{equation} \label{eq:hSYK_second_moment_useful_form}
\mathbb{E} \big[ Z_{\textrm{bSYK}}^2 \big] = \int \textrm{d}Q \, e^{N \beta^2 Q^{2p} - N \Lambda_2[Q]},
\end{equation}
with
\begin{equation} \label{eq:hSYK_second_moment_Q_entropy}
e^{-N \Lambda_2[Q]} \equiv \int \textrm{d}\lambda \, e^{-N \beta^2 \lambda Q} \int \mathcal{D}G \mathcal{D}F \, e^{N A_2[G, F; \lambda]}.
\end{equation}
One can show that
\begin{equation} \label{eq:hSYK_second_moment_Q_entropy_normalization}
\int \textrm{d}Q \, e^{-N \Lambda_2[Q]} = \mathbb{E} \big[ Z_{\textrm{bSYK}} \big] ^2.
\end{equation}
This is easiest to see starting from Eq.~\eqref{eq:hSYK_replicated_path_integral} in Appendix~\ref{app:replicated_action}.
As a result, we can write
\begin{equation} \label{eq:hSYK_variance_useful_form}
\frac{\textrm{Var} \big[ Z_{\textrm{bSYK}} \big] }{\mathbb{E} \big[ Z_{\textrm{bSYK}} \big] ^2} = \frac{\int \textrm{d}Q \left( e^{N \beta^2 Q^{2p}} - 1 \right) e^{-N \Lambda_2[Q]}}{\int \textrm{d}Q \, e^{-N \Lambda_2[Q]}}.
\end{equation}

We have already established that $\Lambda_2'[0] = 0$, by virtue of $Q = 0$ satisfying the saddle-point equations.
Furthermore, evaluating Eq.~\eqref{eq:hSYK_second_moment_Q_entropy} by saddle-point demonstrates that $\Lambda_2[Q]$ is the Legendre-Fenchel transform of $A_2[G_{\textrm{eq}}, F_{\textrm{eq}}; \lambda]$ and must therefore be convex~\cite{Rockafellar1970Convex}.
Thus $Q = 0$ is the \textit{unique} minimum of $\Lambda_2[Q]$.
Finally, a direct calculation starting from Eq.~\eqref{eq:hSYK_second_moment_Q_entropy} gives
\begin{equation} \label{eq:hSYK_second_moment_Q_entropy_curvature}
\Lambda_2''[0] = 2 \left( \int_0^1 \textrm{d}\tau \, G_{\textrm{eq}}(\tau) \right) ^{-2},
\end{equation}
i.e., the curvature of $\Lambda_2$ remains non-zero even as $\beta \rightarrow 0$.

These observations imply that $\beta^2 Q^{2p} - \Lambda_2[Q]$, although not itself concave, has its global maximum at $Q = 0$ for $\beta$ less than a certain non-zero value.
Furthermore, for such $\beta$,
\begin{equation} \label{eq:hSYK_second_moment_correction_order}
\begin{aligned}
& \int \textrm{d}Q \left( e^{N \beta^2 Q^{2p}} - 1 \right) e^{-N \Lambda_2[Q]} \\
& \qquad \qquad = O \left( \frac{1}{N^{p-1}} \right) \cdot \int \textrm{d}Q \, e^{-N \Lambda_2[Q]}.
\end{aligned}
\end{equation}
This establishes what we claimed: there exists a critical temperature above which $\textrm{Var}[Z_{\textrm{bSYK}}] / \mathbb{E} [Z_{\textrm{bSYK}}]^2 \rightarrow 0$ in the thermodynamic limit and $f_{\textrm{bSYK}} = f^{(\textrm{ann})}_{\textrm{bSYK}}$.

\subsection{Quantum p-spin} \label{subsec:quantum_p_spin}

The quantum $p$-spin model is given by Eq.~\eqref{eq:p_spin_Hamiltonian}, reproduced here:
\begin{equation}
H_p =  \sum_{I A} J_I^A \hat{\sigma}_{i_1}^{\alpha_1} \cdots \hat{\sigma}_{i_p}^{\alpha_p} \tag{\ref{eq:p_spin_Hamiltonian}}.
\end{equation}
Our treatment of it will be very similar to that of the bosonic SYK model, and we include it here to highlight the generality of the method.
For that reason, we will present only the results of each step, and leave the details to be filled in by analogy with Sec.~\ref{subsec:hard_core_SYK}.

We express the partition function in terms of spin coherent states $| \Omega \rangle$.
In fact, the only features of the states that we need are the identities~\cite{Lieb1973Classical,Auerbach1994Interacting}
\begin{equation} \label{eq:spin_coherent_state_identity_1}
1 = \int \left( \prod_i \frac{\textrm{d}\Omega_i}{2\pi} | \Omega_i \rangle \langle \Omega_i | \right) ,
\end{equation}
\begin{equation} \label{eq:spin_coherent_state_identity_2}
\hat{S}_{i_1}^{\alpha_1} \cdots \hat{S}_{i_p}^{\alpha_p} = \int \left( \prod_i \frac{\textrm{d}\Omega_i}{2\pi} | \Omega_i \rangle \langle \Omega_i | \right) \left( \frac{3}{2} \right) ^p \Omega_{i_1}^{\alpha_1} \cdots \Omega_{i_p}^{\alpha_p}.
\end{equation}
The integrals are over the unit sphere, and $\Omega_i^x = \sin{\theta_i} \cos{\phi_i}$, etc.
The partition function is
\begin{equation} \label{eq:p_spin_original_partition}
Z_p = \int \prod_i \mathcal{D}\Omega_i \, e^{-\beta \left( \frac{3}{2} \right) ^p \sum_{IA} J_I^A \int_0^1 \textrm{d}\tau \, \Omega_{i_1}^{\alpha_1}(\tau) \cdots \Omega_{i_p}^{\alpha_p}(\tau)}.
\end{equation}
In this notation, we are including the overlaps between coherent states in the integration measure, i.e.,
\begin{equation} \label{eq:p_spin_integration_measure}
\mathcal{D}\Omega_i \equiv \prod_{\tau} \frac{\textrm{d}\Omega_i(\tau)}{2\pi} \langle \Omega_i(\tau + \textrm{d}\tau) | \Omega_i(\tau) \rangle .
\end{equation}
We will never need to express the overlaps in continuum notation.

The replicated, disorder-averaged partition function is
\begin{equation} \label{eq:p_spin_replicated_partition}
\mathbb{E} \big[ Z_p^n \big] = \int \prod_{rr'} \mathcal{D}G_{rr'} \mathcal{D}F_{rr'} \, e^{N \Phi_n[G, F]},
\end{equation}
with
\begin{widetext}
\begin{equation} \label{eq:p_spin_mean_field_action}
\Phi_n[G, F] \equiv \frac{\beta^2}{4} \left( \frac{3}{2} \right) ^{2p} \sum_{rr'} \int_0^1 \textrm{d}\tau \textrm{d}\tau' \Big( G_{rr'}(\tau, \tau')^p - F_{rr'}(\tau, \tau') G_{rr'}(\tau, \tau') \Big) + \ln{\int \prod_r \mathcal{D}\Omega_r \, e^{-S^{(\textrm{eff})}[\Omega]}},
\end{equation}
\begin{equation} \label{eq:p_spin_effective_action}
S^{(\textrm{eff})}[\Omega] \equiv -\frac{\beta^2}{12} \left( \frac{3}{2} \right) ^{2p} \sum_{rr'} \int_0^1 \textrm{d}\tau \textrm{d}\tau' \, F_{rr'}(\tau, \tau') \sum_{\alpha} \Omega_r^{\alpha}(\tau) \Omega_{r'}^{\alpha}(\tau').
\end{equation}
The saddle-point equations are then
\begin{equation} \label{eq:p_spin_G_equation}
F_{rr'}(\tau, \tau') = p \, G_{rr'}(\tau, \tau')^{p-1},
\end{equation}
\begin{equation} \label{eq:p_spin_F_equation}
G_{rr'}(\tau, \tau') = \frac{1}{3} \sum_{\alpha} \big< \Omega_r^{\alpha}(\tau) \Omega_{r'}^{\alpha}(\tau') \big> _{\textrm{eff}}.
\end{equation}
The notation is the same as before: $r$ and $r'$ are replica indices, and $\langle \; \cdot \; \rangle _{\textrm{eff}}$ denotes an expectation value using the effective action of Eq.~\eqref{eq:p_spin_effective_action}.

\subsubsection*{Low temperature}

The $n = 1$ action is
\begin{equation} \label{eq:p_spin_annealed_action}
\Phi_1[G, F] = \frac{\beta^2}{4} \left( \frac{3}{2} \right) ^{2p} \int_0^1 \textrm{d}\tau \Big( G(\tau)^p - F(\tau) G(\tau) \Big) + \ln{\int \mathcal{D}\Omega \, e^{\frac{\beta^2}{12} \left( \frac{3}{2} \right) ^{2p} \int_0^1 \textrm{d}\tau \textrm{d}\tau' F(\tau - \tau') \sum_{\alpha} \Omega^{\alpha}(\tau) \Omega^{\alpha}(\tau')}}.
\end{equation}
We again make a static ansatz for $F(\tau)$, which will turn out to be consistent at low temperature.
A Hubbard-Stratonovich transformation gives
\begin{equation} \label{eq:p_spin_annealed_transformed_action}
\Phi_1[G, F] = \frac{\beta^2}{4} \left( \frac{3}{2} \right) ^{2p} \int_0^1 \textrm{d}\tau \Big( G(\tau)^p - F(\tau) G(\tau) \Big) + \ln{\int \frac{\textrm{d}^3 h}{\sqrt{8 \pi^3}} \, e^{-\frac{1}{2} \sum_{\alpha} h_{\alpha}^2} \int \mathcal{D}\Omega \, e^{\beta \left( \frac{3}{2} \right) ^p \sqrt{\frac{F}{6}} \sum_{\alpha} h_{\alpha} \int_0^1 \textrm{d}\tau \Omega^{\alpha}(\tau)}}.
\end{equation}
The remaining path integral is that of a single spin-1/2 in a magnetic field proportional to $\vec{h}$, thus we can evaluate the $\vec{\Omega}(\tau) \cdot \vec{\Omega}(\tau')$ correlator directly:
\begin{equation} \label{eq:p_spin_annealed_low_T_solution}
G = \frac{1}{27} + \frac{2}{27} \frac{\int \frac{\textrm{d}^3 h}{\sqrt{8 \pi^3}} e^{-\frac{1}{2} |\vec{h}|^2} \cosh{\frac{\beta (1 - 2\tau)}{2} \left( \frac{3}{2} \right) ^{p-1} \sqrt{\frac{F}{6}} |\vec{h}|}}{\int \frac{\textrm{d}^3 h}{\sqrt{8 \pi^3}} e^{-\frac{1}{2} |\vec{h}|^2} \cosh{\frac{\beta}{2} \left( \frac{3}{2} \right) ^{p-1} \sqrt{\frac{F}{6}} |\vec{h}|}},
\end{equation}
which is simply $G \sim 1/27$ in the limit $\beta \rightarrow \infty$ with $\tau \gg 1/\beta$.
Finally, the annealed free energy is
\begin{equation} \label{eq:p_spin_annealed_free_energy_result}
f^{(\textrm{ann})}(\beta) \sim - \frac{\beta}{4} \frac{1}{12^p} + \cdots,
\end{equation}
which indeed diverges at low temperature.

\subsubsection*{High temperature}

Using the same notation as in Sec.~\ref{subsec:hard_core_SYK}, the $n = 2$ action is
\begin{equation} \label{eq:p_spin_second_moment_action}
\begin{aligned}
\Phi_2[G, F; Q, \lambda] =& \, \frac{\beta^2}{2} \left( \frac{3}{2} \right) ^{2p} \int_0^1 \textrm{d}\tau \Big( G(\tau)^p - F(\tau) G(\tau) + Q^p - \lambda Q \Big) \\
+& \ln{\int \mathcal{D}\Omega_1 \mathcal{D}\Omega_2 \, e^{\frac{\beta^2}{12} \left( \frac{3}{2} \right) ^{2p} \int_0^1 \textrm{d}\tau \textrm{d}\tau' F(\tau - \tau') \sum_{\alpha} \big( \Omega_1^{\alpha}(\tau) \Omega_1^{\alpha}(\tau') + \Omega_2^{\alpha}(\tau) \Omega_2^{\alpha}(\tau') \big) }}\\[-6pt]
& \qquad \qquad \qquad \cdot e^{\frac{\beta^2}{12} \left( \frac{3}{2} \right) ^{2p} \lambda \int_0^1 \textrm{d}\tau \textrm{d}\tau' \sum_{\alpha} \big( \Omega_1^{\alpha}(\tau) \Omega_2^{\alpha}(\tau') + \Omega_2^{\alpha}(\tau) \Omega_1^{\alpha}(\tau') \big) }.
\end{aligned}
\end{equation}
One saddle-point of $\Phi_2$ is at
\begin{equation} \label{eq:p_spin_second_moment_uncoupled_saddle}
G(\tau) = G_{\textrm{eq}}(\tau), \quad F(\tau) = F_{\textrm{eq}}(\tau), \quad Q = 0, \quad \lambda = 0,
\end{equation}
\end{widetext}
where $G_{\textrm{eq}}(\tau)$ and $F_{\textrm{eq}}(\tau)$ are the order parameters which maximize $\Phi_1$.
Without any coupling between the two replicas in Eq.~\eqref{eq:p_spin_second_moment_action}, $\langle \Omega_1^{\alpha}(\tau) \Omega_2^{\alpha}(\tau') \rangle = \langle \Omega_1^{\alpha}(\tau) \rangle \langle \Omega_2^{\alpha}(\tau') \rangle$, and $\langle \Omega_r^{\alpha}(\tau) \rangle = 0$ owing to the (statistical) symmetry of the original Hamiltonian.
By establishing that this is the dominant saddle-point at high temperature, we show that $\textrm{Var}[Z_p] / \mathbb{E}[Z_p]^2 \rightarrow 0$ and $f_p = f^{(\textrm{ann})}_p$ with probability 1, exactly as done for the hard-core bosonic model.

Write $\mathbb{E}[Z_p^2]$ as
\begin{equation} \label{eq:p_spin_partition_second_moment_decomposition}
\mathbb{E} \big[ Z_p^2 \big] = \int \textrm{d}Q \, e^{N \frac{\beta^2}{2} \left( \frac{3}{2} \right) ^{2p} Q^p - N \Lambda_2[Q]},
\end{equation}
\begin{equation} \label{eq:p_spin_Q_entropy_decomposition}
e^{-N \Lambda_2[Q]} \equiv \int \textrm{d}\lambda \, e^{-N \frac{\beta^2}{2} \left( \frac{3}{2} \right) ^{2p} \lambda Q} \int \mathcal{D}G \mathcal{D}F \, e^{N A_2[G, F; \lambda]},
\end{equation}
where $A_2$ consists of all the remaining terms in $\Phi_2$.
By the same arguments as in Sec.~\ref{subsec:hard_core_SYK}, we have that: $\Lambda_2'[0] = 0$, $\Lambda_2[Q]$ is convex, and
\begin{equation} \label{eq:p_spin_Q_entropy_curvature}
\Lambda_2''[0] = 3 \left( \int_0^1 \textrm{d}\tau \, G_{\textrm{eq}}(\tau) \right) ^{-2}.
\end{equation}
Thus for $\beta$ less than some non-zero value, $\mathbb{E}[Z_p^2]$ is dominated by $Q = 0$ and
\begin{equation} \label{eq:p_spin_second_moment_scaling}
\textrm{Var} \big[ Z_p \big] \sim O \left( \frac{1}{N^{\frac{p}{2} - 1}} \right) \mathbb{E} \big[ Z_p^2 \big] .
\end{equation}
The free energy then agrees with the annealed value~\footnote{Note that in the $p = 2$ model, $\textrm{Var}[Z_p]/\mathbb{E}[Z_p]^2$ approaches a constant rather than 0. One can show that this constant is less than 1 for $\beta$ less than a critical value, in which case Chebyshev's inequality proves that the free energy equals the annealed value with \textit{some} non-zero probability. This is a slightly weaker result, although if one assumes the free energy is self-averaging (as is standard), then we again have that the annealed approximation is correct with probability 1.}.

\section{The danger in 1/N expansions} \label{sec:N_expansion}

The preceding sections have been in the spirit of the replica formalism, but there is another technique for studying all-to-all disordered models: expanding the free energy in powers of system size $N$.
Many studies of the SYK model and its variants have taken the latter approach~\cite{Maldacena2016Remarks,Gross2017Generalization,Berkooz2019Towards}.
Although in principle one could obtain all of the above results through a $1/N$ expansion, the correct low-temperature physics cannot be identified without taking subtle issues of convergence seriously.
The purpose of this final section is to present an example of the issues which arise, as an argument in favor of the replica method over $1/N$ expansions.

First consider the structure of a $1/N$ expansion, say for the quantum $p$-spin model for concreteness.
Suppose one wishes to compute the moments of the partition function, $\mathbb{E}[Z_p^n]$.
We can expand the exponentials:
\begin{equation} \label{eq:partition_function_moment_expansion}
\begin{aligned}
& \mathbb{E} \big[ \big( \textrm{Tr} e^{-\beta H_p} \big) ^n \big] \\
& \; = \mathbb{E} \left[ \left( \sum_{L=0}^{\infty} \frac{1}{L!} \textrm{Tr} (-\beta H_p)^L \right) ^n \right] \\
& \; = \sum_{L_1, \ldots, L_n = 0}^{\infty} \frac{(-\beta)^{L_1 + \cdots + L_n}}{L_1! \cdots L_n!} \textrm{Tr}_1 \cdots \textrm{Tr}_n \mathbb{E} \big[ H_{p,1}^{L_1} \cdots H_{p,n}^{L_n} \big] .
\end{aligned}
\end{equation}
Note that in the second line, the $n$ replicas are considered as separate degrees of freedom, each with its own trace.
However, the $L_r$ factors of $H_{p,r}$ all involve the same spins of replica $r$, and every factor contains the same Gaussian couplings $J_I^A$.
Since $H_p$ is linear in the couplings, the product $H_{p,1}^{L_1} \cdots H_{p,n}^{L_n}$ is a sum of products of Gaussians, and the disorder average is given by all pairwise contractions according to Wick's theorem.
These features are all naturally expressed in terms of chord diagrams~\cite{Berkooz2018Chord,Berkooz2019Towards}, which we describe in Appendix~\ref{app:chords}.
Evaluation of Eq.~\eqref{eq:partition_function_moment_expansion} is then reduced to a sum over chord diagrams.
Each diagram comes with a power of $N$, which allows the sum to be organized as a $1/N$ expansion.

We further show in Appendix~\ref{app:chords} that, assuming all $L_r \ll N$, the diagrams having contractions between replicas are subleading.
In other words, the disorder average factors to leading order:
\begin{equation} \label{eq:partition_moment_naive_factoring}
\begin{gathered}
\mathbb{E} \big[ H_{p,1}^{L_1} \cdots H_{p,n}^{L_n} \big] \sim \mathbb{E} \big[ H_{p,1}^{L_1} \big] \cdots \mathbb{E} \big[ H_{p,n}^{L_n} \big], \\
\qquad \qquad \qquad \qquad \qquad \qquad ( L_1, \ldots , L_n \ll N ) .
\end{gathered}
\end{equation}
Since in the thermodynamic limit $L_r \ll N$ for any fixed $L_r$, the naive conclusion would be that the entire sum factors, and thus $\mathbb{E}[Z_p^n] \sim \mathbb{E}[Z_p]^n$.
In particular, $\mathbb{E}[Z_p^2] \sim \mathbb{E}[Z_p]^2$, which would imply by Chebyshev's inequality that $Z_p \sim \mathbb{E}[Z_p]$ with high probability.
Yet we proved in Sec.~\eqref{sec:general_breakdown} that $Z_p \not \sim \mathbb{E}[Z_p]$ at low temperature.

The error is in assuming that the dominant terms of Eq.~\eqref{eq:partition_function_moment_expansion} have $L \sim O(1)$.
Since we expect the energy to be extensive, i.e., $H_p \sim O(N)$, the expansion of $e^{-\beta H_p}$ should be dominated by $L \sim O(N)$.
We must at the very least include such $L$ in our evaluation of Eq.~\eqref{eq:partition_function_moment_expansion}.

The non-commutativity of the operators in the quantum model makes it difficult to be any more quantitative.
Thus instead consider the simpler classical model:
\begin{equation} \label{eq:classical_p_spin_definition}
H_{\textrm{cl}} = - \sum_I J_I \sigma_{i_1}^z \cdots \sigma_{i_p}^z,
\end{equation}
where the sum is again over all multi-indices of $p$ spins, and $\textrm{Var}[J_I] = p! / 2N^{p-1}$.
Note that $p = 2$ is precisely the SK model described in Sec.~\ref{sec:general_breakdown} (Eq.~\eqref{eq:SK_model}).

Every statement made above about the quantum $p$-spin model can also be made about the classical model, and in the classical model we can confirm our suspicion that the breakdown of the annealed approximation appears only at $O(N)$'th order in the $1/N$ expansion.
We start with
\begin{equation} \label{eq:classical_partition_variance_expression}
\mathbb{E} \big[ Z_{\textrm{cl}}^2 \big] = \sum_{L_1, L_2 = 0}^{\infty} \frac{(-\beta)^{L_1 + L_2}}{L_1! L_2!} \textrm{Tr}_1 \textrm{Tr}_2 \mathbb{E} \big[ H_{\textrm{cl},1}^{L_1} H_{\textrm{cl},2}^{L_2} \big] .
\end{equation}
A term of Eq.~\eqref{eq:classical_partition_variance_expression} with given $(L_1, L_2)$ contains $L_1 + L_2$ factors of the Gaussian couplings, which are contracted in pairs.
Some contractions will connect spins on the first replica to spins on the second.
By organizing the expansion in terms of the number $L$ of such pairings, as detailed in Appendix~\ref{app:chords}, we have that
\begin{widetext}
\begin{equation} \label{eq:classical_partition_variance_simple}
\begin{aligned}
\mathbb{E} \big[ Z_{\textrm{cl}}^2 \big] =& \; e^{\frac{N \beta^2}{2}} \sum_{L=0}^{\infty} \frac{1}{L!} \left( \frac{\beta^2 p!}{2N^{p-1}} \right) ^L \textrm{Tr}_1 \textrm{Tr}_2 \sum_{I_1 \cdots I_L} \sigma_{I_1 1}^z \sigma_{I_1 2}^z \cdots \sigma_{I_L 1}^z \sigma_{I_L 2}^z \\
=& \; \mathbb{E} \big[ Z_{\textrm{cl}} \big] ^2 \sum_{L=0}^{\infty} \frac{1}{L!} \left( \frac{N \beta^2}{2} \right) ^L \frac{1}{2^{2N}} \textrm{Tr}_1 \textrm{Tr}_2 \left( \frac{1}{N} \sum_i \sigma_{i1}^z \sigma_{i2}^z \right) ^{pL},
\end{aligned}
\end{equation}
\end{widetext}
where in the first line $\sigma_{I_j r}^z \equiv \prod_{i \in I_j} \sigma_{ir}^z$ ($r$ is the replica index), and in the second line the nested sum has been factored.
We also used that $\mathbb{E}[Z_{\textrm{cl}}] = e^{N(\ln{2} + \beta^2 / 4)}$.

To proceed, write the trace as
\begin{equation} \label{eq:classical_trace_eval}
\begin{aligned}
& \frac{1}{2^{2N}} \textrm{Tr}_1 \textrm{Tr}_2 \left( \frac{1}{N} \sum_i \sigma_{i1} \sigma_{i2} \right) ^{pL} \\
& \qquad = \frac{N}{2} \int_{-1}^1 \textrm{d}Q \, \frac{1}{2^N} \binom{N}{N \frac{1 - Q}{2}} Q^{pL} \\
& \qquad \sim \sqrt{\frac{N}{2 \pi}} \int_{-1}^1 \textrm{d}Q \, \frac{Q^{pL}}{\sqrt{1 - Q^2}} \, e^{N s(Q)},
\end{aligned}
\end{equation}
with
\begin{equation} \label{eq:classical_trace_entropy}
s(Q) = -\frac{1 + Q}{2} \ln{(1 + Q)} - \frac{1 - Q}{2} \ln{(1 - Q)}.
\end{equation}
The integral over $Q$ can be evaluated by saddle point, leaving us with a single sum over $L$.

First consider $L \sim O(1)$ with respect to $N$.
The saddle point is at $Q = 1/2$, and the integral goes as $(pL/2 - 1)!! N^{-pL/2}$.
We have that
\begin{equation} \label{eq:classical_partition_variance_order_1}
\frac{\mathbb{E} \big[ Z_{\textrm{cl}}^2 \big]}{\mathbb{E} \big[ Z_{\textrm{cl}} \big] ^2} \sim 1 + \sum_{L=1}^{\infty} \frac{\left( \frac{pL}{2} - 1 \right) !!}{L!} \left( \frac{\beta^2}{2} \right) ^L N^{\left( 1 - \frac{p}{2} \right) L} .
\end{equation}
Every term in the sum over $L$ is subleading in $N$, for any $\beta$ (except if $p = 2$, in which case see~\footnotemark[\value{footnote}]).
This would seem to say that $\mathbb{E} [Z_{\textrm{cl}}^2] \sim \mathbb{E} [Z_{\textrm{cl}}]^2$, even though that cannot possibly be the correct result at low temperature.

\begin{figure}[t]
\centering
\includegraphics[width=1.0\columnwidth]{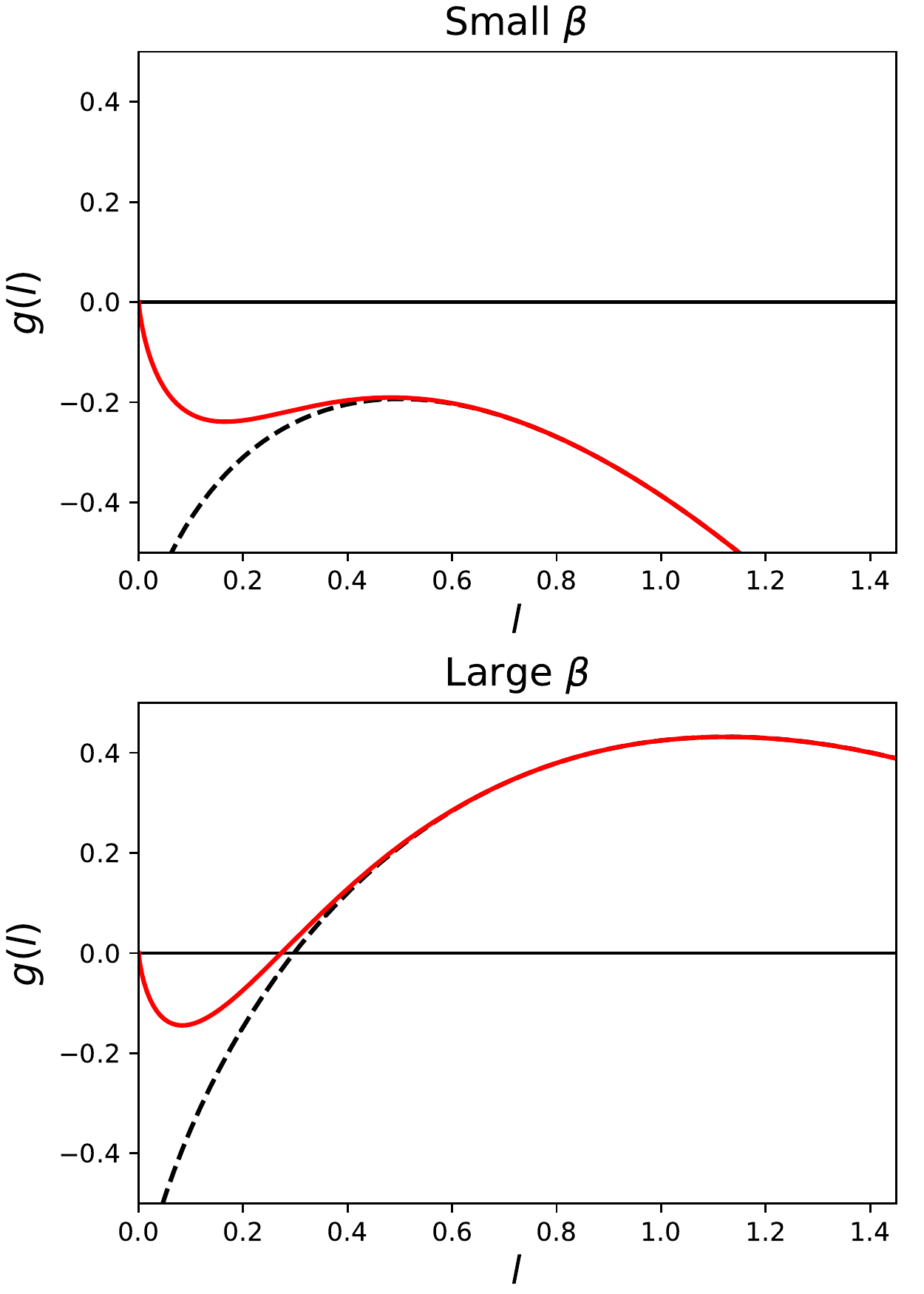}
\caption{The exponent $g(l)$ governing corrections to the annealed free energy, for both small $\beta$ and large $\beta$. Dashed lines are the asymptotic behavior of $g(l)$ at large $l$. The specific parameters used are $p = 6$, small $\beta = 1.0$, large $\beta = 1.5$.}
\label{fig:expansion_exponent}
\end{figure}

However, consider $L \sim O(N)$.
The saddle point $Q^*$ is now given by the equation
\begin{equation} \label{eq:classical_Q_saddle_order_N}
\ln{\frac{1 + Q^*}{1 - Q^*}} = \frac{2pl}{Q^*},
\end{equation}
where $l \equiv L/N$, and the sum over $L$ can be written (ignoring sub-exponential prefactors)
\begin{equation} \label{eq:classical_L_sum_order_N}
\int_0^{\infty} \textrm{d}l \, e^{N \left( l \ln{\frac{e \beta^2}{2l}} + pl \ln{Q^*(l)} + s(Q^*(l)) \right) } \equiv \int_0^{\infty} \textrm{d}l \, e^{N g(l)}.
\end{equation}
This integral is evaluated by saddle point as well.
The limiting behavior of the exponent is
\begin{equation} \label{eq:classical_L_sum_exponent_behaviors}
g(l) \sim \begin{cases} l \ln{\beta^2 l^{\frac{p}{2} - 1}}, \quad & l \ll 1 \\ l \ln{\frac{e \beta^2}{2l}} - \ln{2}, \quad & l \gg 1 \end{cases}.
\end{equation}
See Fig.~\ref{fig:expansion_exponent} as well.
It is clear that if $\beta$ is small, $g(l) < 0$ for all $l > 0$ and thus the integral is $O(1/N)$.
The fluctuations in the partition function are small and the correction to the annealed free energy is indeed subleading.
Yet if $\beta$ is large, the maximum of $g(l)$ is positive.
The fluctuations in $Z_{\textrm{cl}}$ become greater than the mean, and we can no longer claim that the annealed free energy is correct.

We have shown that a $1/N$ expansion of the partition function (and thus of the free energy) converges at small $\beta$ but diverges at large $\beta$.
Furthermore, note that the saddle point of Eq.~\eqref{eq:classical_L_sum_order_N} is at $l^* \sim \beta^2 / 2$ for large $\beta$, i.e., $L^* \sim N \beta^2 / 2$.
Were one to take the $N \rightarrow \infty$ limit before resumming the series, one would miss the divergence entirely, and indeed overlook much of what makes these spin glass models interesting.

\section{Conclusion} \label{sec:conclusion}

We have demonstrated that the annealed approximation breaks down at low temperature in any all-to-all disordered model with finite-body interactions and a tensor product Hilbert space.
This encompasses many in the family of SYK-like models, such as the bosonic variants and the quantum $p$-spin model.
Furthermore, we have shown that, at least in the hard-core bosonic and quantum $p$-spin models (although the technique can easily be generalized), the partition function is self-averaging at high temperature.
Thus we have identified two distinct phases: one in which the free energy equals the annealed value, and one in which it does not.
These results were obtained using rigorous bounds on the annealed free energy and the replica technique.
Note that we did not rely on any of the more cryptic aspects of the replica method (taking the number of replicas to 0 and maximizing rather than minimizing the free energy).
Finally, we have highlighted the subtleties that come with applying $1/N$ expansions to such models.

Strictly speaking, these results are not enough to prove that the models are \textit{spin glasses} at low temperature.
Spin glass order is characterized by an overlap matrix in which the permutation symmetry is broken (``replica symmetry breaking''), whereas we have shown only that the matrix cannot be diagonal.
In more physical terms, a spin glass has multiple low temperature states, whereas we have shown only the existence of \textit{some} low temperature state distinct from the high temperature state.

That said, the results established here do force one to confront the issue of glassiness.
The standard annealed approximation cannot accurately describe the effects of disorder in any tensor product model, and one must use an approach which allows for non-diagonal and potentially symmetry-broken replica order parameters.
In particular, this statement applies to many models of current interest in the context of SYK physics.
Whether replica symmetry is broken or merely non-diagonal in any specific model is an interesting open question which requires further analysis.

As for the relevance of these models to holography, it is still possible that some might have gravitational duals despite the breakdown of the annealed approximation.
The precise dynamics cannot be exactly as in fermionic SYK, since that model is described by the annealed approximation, but a more complex gravitational theory is not ruled out.
It is also possible that glassiness and gravitational dynamics can coexist in an interesting way, e.g., Refs.~\cite{Kachru_2010,anninos2013holographic,Facoetti2019Classical}.
These are all important questions that remain to be investigated.

There is one potential way for the annealed free energy to remain accurate at low temperature even in tensor product models: have an interaction degree which increases with system size.
Note that every bound obtained here no longer diverges if the $p \rightarrow \infty$ limit is taken before the $T \rightarrow 0$ limit.
This does not prove that the annealed approximation holds, but we cannot claim that it must break down in such models.
One example is the double-scaling limit studied in Refs.~\cite{Cotler_2017,Erdos_2014,Berkooz2018Chord,Berkooz2019Towards}, where $p \sim \sqrt{N}$. It was argued that the quantum $p$-spin model has a Schwarzian density of states in this limit.
In view of our results, it would clearly be desirable to have a more detailed understanding of the low-energy physics for general $p$.

Finally, it is interesting to note that every system currently known to have a simple gravitational dual includes fermionic degrees of freedom.
This could be a streetlight effect, perhaps related to the difficulty of reliably studying non-supersymmetric theories at strong coupling.
However, here we have uncovered a general result preventing a wide class of bosonic theories from exhibiting the simplest kind of gravitational dynamics known to occur in a corresponding fermionic theory.
Perhaps this is one example of a general class of constraints which places purely bosonic theories of gravity into the swampland~\cite{vafa2005string}.

\section{Acknowledgements}

We would like to thank C. R. Laumann and A. Kamenev for useful discussions.
This research was performed while CLB held an NRC Research Associateship award at the National Institute of Standards and Technology.

\bibliography{Quenched_Annealed_Biblio}

\appendix

\begin{widetext}

\section{Calculating the replicated action} \label{app:replicated_action}

We present the details for the hard-core bosonic model.
The quantum $p$-spin model and others proceed analogously.

Beginning from Eq.~\eqref{eq:hSYK_partition_path_integral}, we have that
\begin{equation} \label{eq:averaged_hSYK_partition_function}
\begin{aligned}
\mathbb{E} \big[ Z_{\textrm{bSYK}}^n \big] =& \int \prod_{ir} \mathcal{D}h_{ir} \mathcal{D}a_{ir} \mathcal{D}\mu_{ir} \, e^{-\sum_{ir} S^{(0)}[h_{ir}, a_{ir}, \mu_{ir}]} \, \mathbb{E} \left[ e^{-\sum_{II'} \sum_r J_{II'} S_{\textrm{int}}[h_{II'r}, a_{II'r}]} \right] \\
=& \int \prod_{ir} \mathcal{D}h_{ir} \mathcal{D}a_{ir} \mathcal{D}\mu_{ir} \, e^{-\sum_{ir} S^{(0)}[h_{ir}, a_{ir}, \mu_{ir}] + \frac{(p!)^2}{2N^{2p-1}} \sum_{rr'} \sum_{II'} S_{\textrm{int}}[h_{II'r}, a_{II'r}] S_{\textrm{int}}[h_{II'r'}, a_{II'r'}]^*}.
\end{aligned}
\end{equation}
The sums over spins/multi-indices now come alongside sums over replica indices $r, r' \in \{ 1, \cdots, n \}$.
Note that, to leading order in $N$,
\begin{equation} \label{eq:hSYK_action_factoring}
\begin{aligned}
& \frac{(p!)^2}{2N^{2p-1}} \sum_{rr'} \sum_{II'} S_{\textrm{int}}[h_{II'r}, a_{II'r}] S_{\textrm{int}}[h_{II'r'}, a_{II'r'}]^* \\
& \quad = \frac{\beta^2 (p!)^2}{2N^{2p-1}} \sum_{rr'} \int_0^1 \textrm{d}\tau \textrm{d}\tau' \sum_{i_1 < \cdots < i_p} \sum_{i'_1 < \cdots < i'_p} \Big[ a_{i_1 r}(\tau)^* h_{i_1 r}(\tau) \cdots a_{i_p r}(\tau)^* h_{i_p r}(\tau) h_{i'_p r}(\tau)^* a_{i'_p r}(\tau) \cdots h_{i'_1 r}(\tau)^* a_{i'_1 r}(\tau) \\
& \qquad \qquad \qquad \qquad \qquad \qquad \qquad \cdot a_{i'_1 r'}(\tau')^* h_{i'_1 r'}(\tau') \cdots a_{i'_p r'}(\tau')^* h_{i'_p r'}(\tau') h_{i_p r'}(\tau')^* a_{i_p r'}(\tau') \cdots h_{i_1 r'}(\tau')^* a_{i_1 r'}(\tau') \Big] \\
& \quad = \frac{\beta^2}{2N^{2p-1}} \sum_{bb'} \int_0^1 \textrm{d}\tau \textrm{d}\tau' \sum_{i_1 \neq \cdots \neq i_p} \sum_{i'_1 \neq \cdots \neq i'_p} \Big[ a_{i_1 r}(\tau)^* h_{i_1 r}(\tau) h_{i_1 r'}(\tau')^* a_{i_1 r'}(\tau') \cdots a_{i_p r}(\tau)^* h_{i_p r}(\tau) h_{i_p r'}(\tau')^* a_{i_p r'}(\tau') \\
& \qquad \qquad \qquad \qquad \qquad \qquad \qquad \qquad \qquad \; \; \cdot h_{i'_p r}(\tau)^* a_{i'_p r}(\tau) a_{i'_p r'}(\tau')^* h_{i'_p r'}(\tau') \cdots h_{i'_1 r}(\tau)^* a_{i'_1 r}(\tau) a_{i'_1 r'}(\tau')^* h_{i'_1 r'}(\tau') \Big] \\
& \quad \sim \frac{N \beta^2}{2} \sum_{bb'} \int_0^1 \textrm{d}\tau \textrm{d}\tau' \left| \frac{1}{N} \sum_i h_{ir}(\tau)^* a_{ir}(\tau) a_{ir'}(\tau')^* h_{ir'}(\tau') \right| ^{2p}.
\end{aligned}
\end{equation}
Thus the action for $\mathbb{E} [ Z_{\textrm{bSYK}}^n ]$ (not including $S^{(0)}$) is a functional solely of
\begin{equation} \label{eq:hSYK_order_parameter_definition}
G_{rr'}(\tau, \tau') \equiv \frac{1}{N} \sum_i h_{ir}(\tau)^* a_{ir}(\tau) a_{ir'}(\tau')^* h_{ir'}(\tau').
\end{equation}
We write this explicitly inside the path integral by introducing a $\delta$-functional:
\begin{equation} \label{eq:hSYK_replicated_path_integral}
\begin{aligned}
\mathbb{E} \big[ Z_{\textrm{bSYK}}^n \big] &= \int \prod_{rr'} \mathcal{D}G_{rr'} \, e^{\frac{N \beta^2}{2} \sum_{rr'} \int_0^1 \textrm{d}\tau \textrm{d}\tau' G_{rr'}(\tau, \tau')^p G_{r'r}(\tau', \tau)^p} \\
& \cdot \int \prod_{ir} \mathcal{D}h_{ir} \mathcal{D}a_{ir} \mathcal{D}\mu_{ir} \, e^{-\sum_{ir} S^{(0)}[h_{ir}, a_{ir}, \mu_{ir}]} \delta \left( G_{rr'}(\tau, \tau') - \frac{1}{N} \sum_i h_{ir}(\tau)^* a_{ir}(\tau) a_{ir'}(\tau')^* h_{ir'}(\tau') \right) .
\end{aligned}
\end{equation}

In the large-$N$ limit, the path integral is dominated by a specific value of $G_{rr'}(\tau, \tau')$.
We determine this saddle point by introducing a Lagrange multipler $F_{rr'}(\tau, \tau')$, and thus have
\begin{equation} \label{eq:hSYK_canonical_replicated_path_integral}
\begin{aligned}
\mathbb{E} \big[ Z_{\textrm{bSYK}}^n \big] =& \int \prod_{rr'} \mathcal{D}G_{rr'} \mathcal{D}F_{rr'} \, e^{\frac{N \beta^2}{2} \sum_{rr'} \int_0^1 \textrm{d}\tau \textrm{d}\tau' \big( G_{rr'}(\tau, \tau')^p G_{r'r}(\tau', \tau)^p - F_{rr'}(\tau, \tau') G_{r'r}(\tau', \tau) \big) } \\
& \qquad \qquad \cdot \int \prod_{ir} \mathcal{D}h_{ir} \mathcal{D}a_{ir} \mathcal{D}\mu_{ir} \, e^{-\sum_i S^{(\textrm{eff})}[h_i, a_i, \mu_i]},
\end{aligned}
\end{equation}
where $S^{(\textrm{eff})}$ is given by Eq.~\eqref{eq:hSYK_single_site_effective_action}.
The path integral over $h$, $a$, and $\mu$ now factors among the $N$ different sites $i$, and we obtain Eqs.~\eqref{eq:hSYK_integrated_path_integral} and~\eqref{eq:hSYK_derived_order_parameter_action}:
\begin{equation} \tag{\ref{eq:hSYK_integrated_path_integral}}
\mathbb{E} \big[ Z_{\textrm{bSYK}}^n \big] = \int \prod_{rr'} \mathcal{D}G_{rr'} \mathcal{D}F_{rr'} \, e^{N \Phi_n[G, F] },
\end{equation}
\begin{equation} \tag{\ref{eq:hSYK_derived_order_parameter_action}}
\Phi_n[G, F] \equiv \frac{\beta^2}{2} \sum_{rr'} \int_0^1 \textrm{d}\tau \textrm{d}\tau' \Big( G_{rr'}(\tau, \tau')^p G_{r'r}(\tau', \tau)^p - F_{rr'}(\tau, \tau') G_{r'r}(\tau', \tau) \Big) + \ln{\int \prod_r \mathcal{D}h_r \mathcal{D}a_r \mathcal{D}\mu_r e^{-S^{(\textrm{eff})}[h, a, \mu]}}.
\end{equation}
Note that the remaining integration over $h$, $a$, and $\mu$ is for a \textit{single} site.
In return, the action for that site has couplings between different replicas and times.

\section{Chord diagrams} \label{app:chords}

\begin{figure}[t]
\centering
\includegraphics[width=0.5\textwidth]{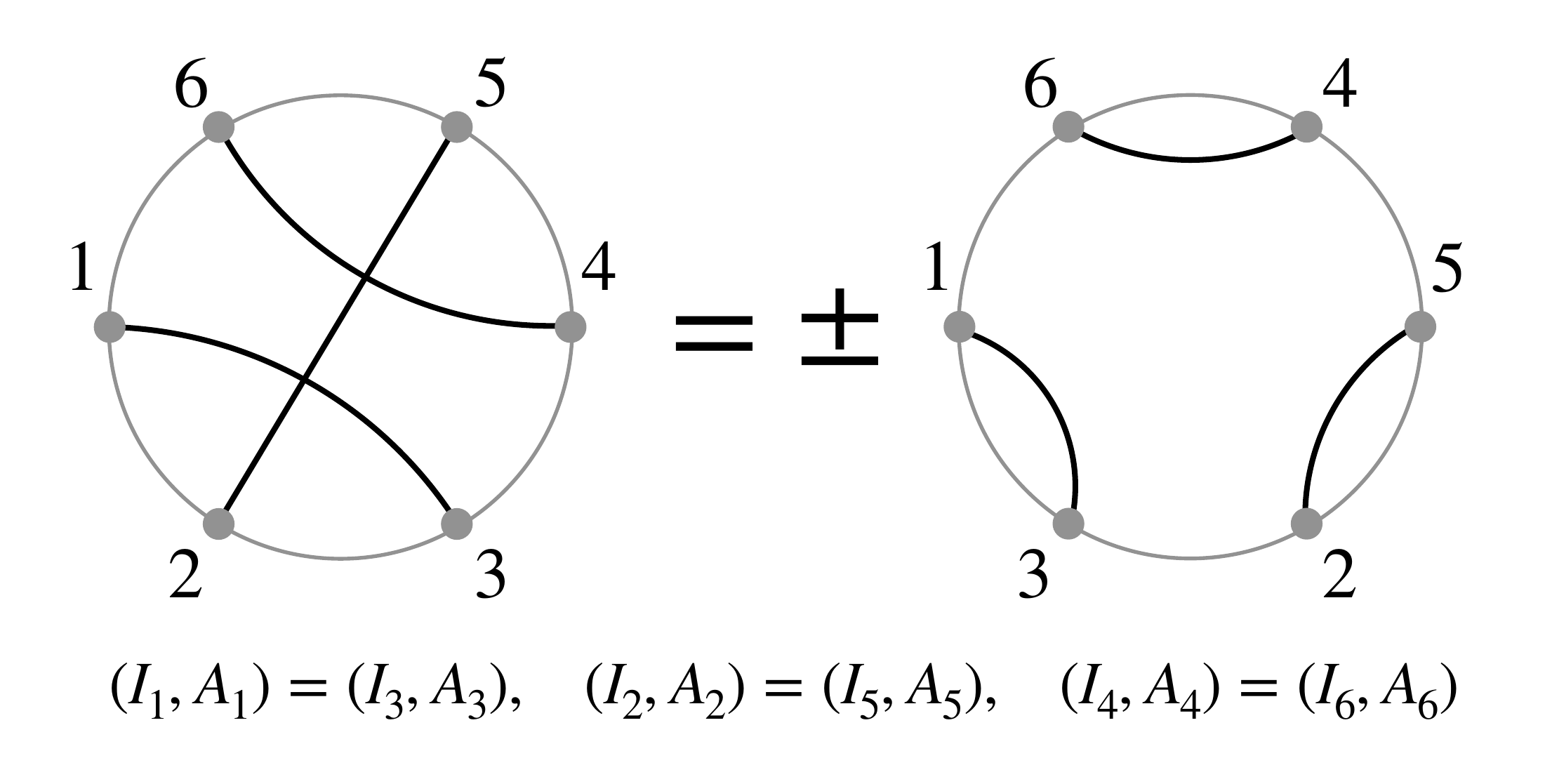}
\caption{(Left) An example of a chord diagram. Each gray marker $l$ represents an operator $\hat{O}_{I_l}^{A_l}$, and each black line indicates that the multi-indices of the two connected markers are equal. (Right) Rearrangement of the operators in the diagram, using that any two operators either commute or anticommute. The amplitude of the rearranged diagram equals that of the original, up to a minus sign which depends on the assigned multi-indices.}
\label{fig:chord_first_moment}
\end{figure}

Starting with the quantum $p$-spin Hamiltonian,
\begin{equation} \label{eq:p_spin_Hamiltonian_compact}
H_p =  \sum_{I A} J_I^A \hat{\sigma}_{i_1}^{\alpha_1} \cdots \hat{\sigma}_{i_p}^{\alpha_p} \equiv \sum_{IA} J_I^A \hat{O}_I^A,
\end{equation}
where we defined $\hat{O}_I^A \equiv \hat{\sigma}_{i_1}^{\alpha_1} \cdots \hat{\sigma}_{i_p}^{\alpha_p}$ for convenience, consider first the problem of calculating the average of the partition function:
\begin{equation} \label{eq:chords_first_moment_expression}
\mathbb{E} \big[ Z_p \big] = \mathbb{E} \big[ \text{Tr} \, e^{-\beta H_p} \big].
\end{equation}
Expanding the exponential and using Eq.~\eqref{eq:p_spin_Hamiltonian}, we have
\begin{equation} \label{eq:chords_first_moment_full_expansion}
\mathbb{E} \big[ Z_p \big] = \sum_{k=0}^{\infty} \frac{(-\beta)^{2k}}{(2k)!} \sum_{I_1 A_1 \cdots I_{2k} A_{2k}} \mathbb{E} \big[ J_{I_1}^{A_1} \cdots J_{I_{2k}}^{A_{2k}} \big] \textrm{Tr} \big[ \hat{O}_{I_1}^{A_1} \cdots \hat{O}_{I_{2k}}^{A_{2k}} \big] .
\end{equation}
Note that all odd-order terms automatically vanished due to the disorder average.

Since the couplings are Gaussian, the $(I_j, A_j)$ multi-indices must be paired up.
Thus the operators $\hat{O}_I^A$ also occur in pairs.
We represent the pairings diagramatically by drawing a circle with $2k$ marked points, one for each insertion of an operator, and drawing $k$ chords through the circle to connect the points in pairs.
Each such picture is called a ``chord diagram'' (see Fig.~\ref{fig:chord_first_moment}).
Finally, we assign a multi-index $(I, A)$ to each chord and sum over all possible assignments.

The simplest chord diagram has $(I_1, A_1) = (I_2, A_2)$, $(I_3, A_3) = (I_4, A_4)$, etc.
In fact, since $(\hat{O}_I^A)^2 = 1$ for all $(I, A)$, evaluating the diagram is trivial: we obtain
\begin{equation} \label{eq:chords_reference_diagram_value}
\frac{(-\beta)^{2k}}{(2k)!} \cdot \left( 3^p \binom{N}{p} \right)^k \cdot \left( \frac{p!}{6(3N)^{p-1}}\right)^{k} \cdot 2^N \sim \frac{1}{(2k)!} \left( \frac{N \beta^2}{2} \right) ^k 2^N.
\end{equation}
The first factor on the left-hand side was explicit in Eq.~\eqref{eq:chords_first_moment_full_expansion}, the second comes from the sum over multi-indices, the third is the variance of $J_I^A$, and the fourth comes simply from tracing over the Hilbert space.
Taking the large-$N$ limit gives the right-hand side.

\textit{If} all the operators were to commute with each other, then every chord diagram would give the same contribution: just rearrange the operators until the members of each pair are adjacent.
Since the total number of pairings is $(2k)!/2^k k!$, the sum over all chord diagrams would give
\begin{equation} \label{eq:chords_first_moment_evaluation}
\begin{aligned}
\mathbb{E} \big[ Z_{p, \textrm{commuting}} \big] =& \; 2^N \sum_{k = 0}^{\infty} \frac{1}{k!} \left( \frac{N \beta^2}{4} \right) ^k \\
=& \; e^{N \left( \ln{2} + \frac{\beta^2}{4} \right) }.
\end{aligned}
\end{equation}
Of course, the operators do not all commute, and $\mathbb{E}[Z_p]$ does not have so simple an expression.
Instead, some choices of $\hat{O}_{I_j}^{A_j}$ and $\hat{O}_{I_l}^{A_l}$ anticommute, e.g., if $\hat{O}_{I_j}^{A_j}$ has a factor of $\hat{\sigma}_1^x$ and $\hat{O}_{I_l}^{A_l}$ has a factor of $\hat{\sigma}_1^y$.
Thus, depending on the specific multi-indices assigned to a diagram, we may acquire a factor of $-1$ in rearranging the operators.
An example is shown in Fig.~\ref{fig:chord_first_moment}.
In place of Eq.~\eqref{eq:chords_reference_diagram_value}, we have
\begin{equation} \label{eq:chords_other_diagram_value}
\frac{(-\beta)^{2k}}{(2k)!} \cdot \left( \sum_{\{ IA \}} \eta_{\{ IA \}} \right) \cdot \left( \frac{p!}{6 (3N)^{p-1}} \right) \cdot 2^N,
\end{equation}
where the sum is over sets of $k$ multi-indices and $\eta_{\{ IA \}}$ is either $1$ or $-1$ depending on the specific multi-indices in question.

\begin{figure}[t]
\centering
\includegraphics[width=0.5\textwidth]{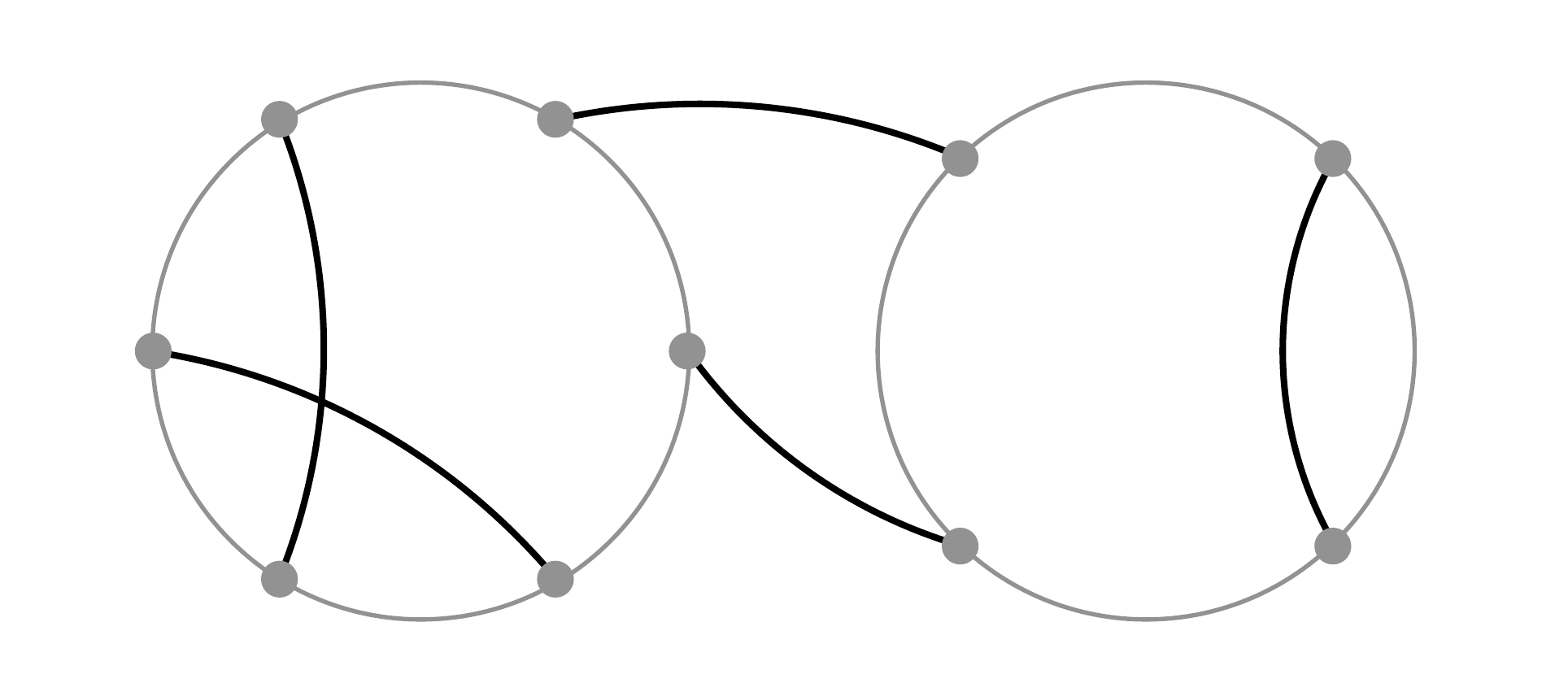}
\caption{An example of a chord diagram relevant for the second moment of $Z_p$.}
\label{fig:chord_second_moment}
\end{figure}

Note that the effect of non-commuting operators is only appreciable in higher-order terms.
Namely, for $k \sim O(1)$ with respect to $N$, Eq.~\eqref{eq:chords_other_diagram_value} gives the same value as Eq.~\eqref{eq:chords_reference_diagram_value}: when choosing $k$ multi-indices involving $p$ spins each, all but an $O(1/N)$ fraction of the possible choices have every spin distinct (and thus commuting).

Rather than dwell further on $\mathbb{E}[Z_p]$, let us turn to the second moment.
Again expanding $e^{-\beta H_p}$, of which there are now two factors, we have
\begin{equation} \label{eq:chords_second_moment_expansion}
\mathbb{E} \big[ Z_p^2 \big] = \sum_{l_1, l_2 = 0}^{\infty} \frac{(-\beta)^{l_1 + l_2}}{l_1! l_2!} \sum_{I_1 A_1 \cdots I_{l_1 + l_2} A_{l_1 + l_2}} \mathbb{E} \big[ J_{I_1}^{A_1} \cdots J_{I_{l_1 + l_2}}^{A_{l_1 + l_2}} \big] \textrm{Tr} \big[ \hat{O}_{I_1}^{A_1} \cdots \hat{O}_{I_{l_1}}^{A_{l_1}} \big] \, \textrm{Tr} \big[ \hat{O}_{I_{l_1 + 1}}^{A_{l_1 + 1}} \cdots \hat{O}_{I_{l_1 + l_2}}^{A_{l_1 + l_2}} \big] .
\end{equation}
In words, there are $l_1 + l_2$ insertions of operators, $l_1$ of which are in the first trace and $l_2$ of which are in the second.
We represent these by two circles with $l_1$ and $l_2$ marked points, respectively.
However, all $l_1 + l_2$ couplings are within the same disorder average, meaning that points can be paired between the circles (see Fig.~\ref{fig:chord_second_moment}).
Organize the sum in Eq.~\eqref{eq:chords_second_moment_expansion} by the number $l$ of pairs connecting the two:
\begin{equation} \label{eq:chords_second_moment_organized}
\mathbb{E} \big[ Z_p^2 \big] = \sum_{l=0}^{\infty} \sum_{k_1, k_2 = 0}^{\infty} \frac{(-\beta)^{2l + 2k_1 + 2k_2}}{(l + 2k_1)! (l + 2k_2)!} \left( \frac{p!}{6(3N)^{p-1}} \right) ^{l + k_1 + k_2} \sum_{\{ IA \}} \eta_{\{ IA \}}.
\end{equation}
The inner sum is over sets of $l + k_1 + k_2$ multi-indices such that the two traces are both non-vanishing, and $\eta_{\{ IA \}}$ is again either $1$ or $-1$ depending on the specific multi-indices.

Note that for the $l = 0$ terms, the $\{ IA \}$ sum factors into two separate sums, one for each circle.
Furthermore, the sums over $k_1$ and $k_2$ are then precisely those that gave us $\mathbb{E}[Z_p]$.
Thus
\begin{equation} \label{eq:chords_variance_expansion}
\textrm{Var} \big[ Z_p \big] = \sum_{l=1}^{\infty} \sum_{k_1, k_2 = 0}^{\infty} \frac{(-\beta)^{2l + 2k_1 + 2k_2}}{(l + 2k_1)! (l + 2k_2)!} \left( \frac{p!}{6(3N)^{p-1}} \right) ^{l + k_1 + k_2} \sum_{\{ IA \}} \eta_{\{ IA \}}.
\end{equation}

For $l = 1$, $\eta_{\{ IA \}} = 0$.
This is because each $\hat{O}_I^A$ is traceless, and the two operators paired between the circles are each left unpaired in their respective traces.

For $l = 2$, let us count the powers of $N$.
Assume that $k_1, k_2 \sim O(1)$ as well, so that we can ignore minus signs as discussed above.
Yet we do still need to ensure that every factor of $\hat{\sigma}_i^{\alpha}$ occurs in pairs to survive the trace.
This restricts the number of sums over spin indices in $\sum_{\{ IA \}}$ to $pk_1 + pk_2 + p$: each of the multi-indices within each circle can be summed freely, but the two which connect the circles must have every index paired with each other.
The counting for other $l \sim O(1)$ is analogous, and once we include the number of contractions, we have
\begin{equation} \label{eq:chords_variance_low_order_scaling}
\begin{aligned}
& \frac{(-\beta)^{2l + 2k_1 + 2k_2}}{(l + 2k_1)! (l + 2k_2)!} \left( \frac{p!}{6(3N)^{p-1}} \right) ^{l + k_1 + k_2} \sum_{\{ IA \}} \eta_{\{ IA \}} \\
& \qquad \qquad \sim \frac{(-\beta)^{2l + 2k_1 + 2k_2}}{(2k_1)! (2k_2)! \, l!} \left( \frac{p!}{6(3N)^{p-1}} \right) ^{l + k_1 + k_2} \frac{(3N)^{pk_1 + pk_2 + \frac{lp}{2}}}{p!^{l + k_1 + k_2}} \\
& \qquad \qquad = \frac{1}{(2k_1)! (2k_2)! \, l!} \left( \frac{N \beta^2}{2} \right) ^{k_1 + k_2} \left( \frac{\beta^2}{2 \cdot 3^{\frac{p}{2}}} \right) ^l N^{-\left( \frac{p}{2} - 1 \right) l}.
\end{aligned}
\end{equation}

Note that, at least for $p > 2$, all $l \neq 0$ are suppressed by powers of $N$ relative to $l = 0$.
If we were to naively sum this expression over all $(l, k_1, k_2)$, we would be led to believe that $\textrm{Var}[Z_p] / \mathbb{E}[Z_p]^2 \rightarrow 0$ as $N \rightarrow \infty$, regardless of $\beta$.
Yet we have proven in the main text that this cannot be true.
The resolution, as also discussed in the main text, is that Eq.~\eqref{eq:chords_variance_low_order_scaling} holds only for $l, k_1, k_2 \sim O(1)$, whereas we need to sum over all values \textit{at fixed} $N$.
Once $(l, k_1, k_2)$ become comparable to $N$, not only do anticommuting operators begin to matter, but the combinatorics of the chord diagrams changes.
This second point is demonstrated explicitly in the main text.

\end{widetext}

\end{document}